\RequirePackage{fix-cm}
\documentclass[smallextended]{svjour3}       

\usepackage{amsmath}
\usepackage{amssymb}
\usepackage{lineno}

\usepackage{graphicx}
\usepackage{epstopdf}
\usepackage[round]{natbib}
\usepackage{color, cancel}

\makeatletter
\renewcommand{\@biblabel}[1]{\quad#1.}

\makeatother

\renewcommand{\figurename}{Fig.}

\begin{document}
\bibliographystyle{spbasic}

\title{Spatial memory and taxis-driven pattern formation in model ecosystems
}

\titlerunning{Diffusion-taxis systems}        

\author{Jonathan R. Potts         \and
        Mark A. Lewis 
}

\institute{Jonathan R. Potts \at
              School of Mathematics and Statistics, University of Sheffield, Hicks Building, Hounsfield Road, Sheffield, S3 7RH, UK \\
              Tel.: +44-222-3729\\
              \email{j.potts@sheffield.ac.uk}           
           \and
           Mark A. Lewis \at
               Departments of Mathematical and Statistical Sciences and Biological Sciences, CAB632, University of Alberta, Edmonton, Alberta, T6G 2G1, Canada
}

\date{Received: date / Accepted: date}

\maketitle

\begin{abstract}
Mathematical models of spatial population dynamics typically focus on the interplay between dispersal events and birth/death processes.  However, for many animal communities, significant arrangement in space can occur on shorter timescales, where births and deaths are negligible.  This phenomenon is particularly prevalent in populations of larger, vertebrate animals who often reproduce only once per year or less.  To understand spatial arrangements of animal communities on such timescales, we use a class of diffusion-taxis equations for modelling inter-population movement responses between $N \geq 2$ populations.  These systems of equations incorporate the effect on animal movement of both the current presence of other populations and the memory of past presence encoded either in the environment or in the minds of animals.  We give general criteria for the spontaneous formation of both stationary and oscillatory patterns, via linear pattern formation analysis.  For $N=2$, we classify completely the pattern formation properties using a combination of linear analysis and non-linear energy functionals.  In this case, the only patterns that can occur asymptotically in time are stationary.  However, for $N \geq 3$, oscillatory patterns can occur asymptotically, giving rise to a sequence of period-doubling bifurcations leading to patterns with no obvious regularity, a hallmark of chaos.  
Our study highlights the importance of understanding between-population animal movement for understanding spatial species distributions, something that is typically ignored in species distribution modelling, and so develops a new paradigm for spatial population dynamics. 
\end{abstract}

\section{Introduction}

Mathematical modelling of spatial population dynamics has a long history of uncovering the mechanisms behind a variety of observed patterns, from predator-prey interactions \citep{pascual1993, lugomckane2008,sunetal2012} to biological invasions \citep{petrovskiietal2002, hastings2005spatial, lewisetal2016} to inter-species competition \citep{hastings1980, durrettlevin1994, girardinnadin2015}.  These models typically start with a mathematical description of the birth and death processes, then add spatial aspects in the form of dispersal movements.  Such movements are often assumed to be diffusive \citep{okubolevin2013}, but sometimes incorporate elements of taxis \citep{kareivaodell1987, leeetal2009, pottspetrovskii2017}.  The resulting models are often systems of reaction-advection-diffusion (RAD) equations, which are amenable to pattern-formation analysis via a number of established mathematical techniques \citep{murray2003}.

An implicit assumption behind these RAD approaches is that the movement processes (advection and diffusion) take place on the same temporal scale as the birth and death processes (reaction).    However, many organisms will undergo significant movement over much shorter time-scales.  For example, many larger animals (e.g. most birds, mammals, and reptiles) will reproduce only once per year, but may rearrange themselves in space quite considerably in the intervening period between natal events.  These rearrangements can give rise to emergent phenomena such as the `landscape of fear' \citep{laundreetal2010}, aggregations of co-existent species \citep{murrelllaw2003}, territoriality \citep{pottslewis2014}, home ranges \citep{briscoeetal2002, borgeretal2008}, and spatial segregation of interacting species \citep{shigesada1979spatial}.

Indeed, the study of organism movements has led, in the past decade or two, to the emergence of a whole subfield of ecology, dubbed `movement ecology' \citep{nathanetal2008, nathangiuggioli2013}.  This is gaining increasing attention by both statisticians \citep{hootenetal2017} and empirical ecologists \citep{kaysetal2015, haysetal2016}, in part driven by recent rapid technological advances in biologging \citep{williamsetalinrev}.  Often, a stated reason for studying movement is to gain insight into space-use patterns \citep{vanaketal2013, avgaretal2015, flemingetal2015, avgaretal2016}.  Yet despite this, we lack a good understanding of the spatial pattern-formation properties of animal movement models over time-scales where birth and death effects are minimal.

To help rectify this situation, we introduce here a class of models that focuses on one particular type of movement: taxis of a population in response to the current or recent presence of foreign populations.  This covers several ideas within the ecological literature.  One is the movement of a species away from areas where predator or competitor species reside, often dubbed the `landscape of fear' \citep{laundreetal2010,gallagheretal2017}.  The opposing phenomenon is that of predators moving towards prey, encapsulated in prey-taxis models \citep{kareivaodell1987, leeetal2009}.  Many species exhibit mutual avoidance, which can be either inter-species avoidance or intra-species avoidance.  The latter gives rise to territoriality and there is an established history of modelling efforts devoted to its study \citep{adams2001, moorcroftlewis2006, pottslewis2014}.  Likewise, some species exhibit mutual attraction due to benefits of co-existence \citep{murrelllaw2003,kneitelchase2004,vanaketal2013}.  Since some of these phenomena are inter-specific and others are intra-specific, we use the word `population' to mean a group of organisms that are all modelled using the same equation, noting once and for all that `population' may be used to mean an entire species (for modelling inter-species interactions, e.g. the landscape of fear), or it may refer to a group within a single species (for intra-species interactions, e.g. territoriality).

There are various processes by which one population can sense the presence of others.  One is by directly sensing organism presence by sight or touch.  However, it is perhaps more common for the presence of others to be advertised indirectly.  This could either be due to marks left in the landscape, a process sometimes known as stigmergy \citep{giuggiolietal2013}, or due to memory of past interactions \citep{faganetal2013, pottslewis2016a}.  We show here that these three interaction processes (direct, stigmergic, memory) can all be subsumed under a single modelling framework.

The resulting model is a system of $N$ diffusion-taxis equations, one for each of $N$ populations.  We analyse this system using a combination of linear pattern formation analysis \citep{turing1952}, energy functionals (non-linear), and numerical bifurcation analysis.
  We classify completely the pattern formation properties for $N=2$, noting that here only stationary patterns can form.  For $N=3$, we show that, as well as there being parameter regimes where stationary patterns emerge, oscillatory patterns can emerge for certain parameter values, where patterns remain transient and never settle to a steady state.  In these regimes, we observe both periodic behaviour and behaviour where the period is much less regular.  These irregular regimes emerge through a sequence of period-doubling bifurcations, a phenomenon often associated with the emergence of chaos.  

The fact that inter-population taxis processes can give rise to perpetually changing, possibly chaotic, spatial patterns is a key insight into the study of species distributions.  Researchers often look to explain such transient spatial patterns by examining changes in the underlying environment.  However, we show that continually changing patterns can emerge without the need to impose any environmental effect.  As such, our study highlights the importance of understanding inter-population movement responses for gaining a full understanding of the spatial distribution of ecological communities, and helps link movement ecology to population dynamics in a non-speculative way.

\section{The modelling framework}

Our general modelling framework considers $N$ populations, each of which has a fixed overall size.  For each population, the constituent individuals move in space through a combination of a diffusive process and a tendency to move towards more attractive areas and away from those that are less attractive.  Denoting by $u_i({\bf x},t)$ the probability density function of population $i$ at time $t$ ($i \in \{1,\dots,N\}$), and by $A_i({\bf x},t)$ the attractiveness of location ${\bf x}$ to members of population $i$ at time $t$, we construct the following movement model
\begin{linenomath*}\begin{align}
\label{gen_model}
\frac{\partial u_i}{\partial t} = D_i\nabla^2 u_i - c_{i} \nabla \cdot\left(u_i \nabla A_i \right),
\end{align}\end{linenomath*}
where $D_i>0$ is the magnitude of the diffusive movement of population $i$ and $c_i\geq 0$ is the magnitude of the drift tendency towards more attractive parts of the landscape.  

Here, we assume that the attractiveness of a point ${\bf x}$ on the landscape at time $t$ is determined by the presence of individuals from other populations.  We look at three scenarios.  For some organisms, particularly very small ones such as amoeba, there may be sufficiently many individuals constituting each population so that the probability density function is an accurate descriptor of the number of individuals present at each part of space.  This is Scenario 1.  In this case, the attractiveness of a part of space to population $i$ may simply be proportional to the weighted sum of the probability density functions of all the other populations, or possibly a locally-averaged probability density.  In other words
\begin{linenomath*}\begin{align}
\label{Aisimp}
\mbox{\bf Scenario 1: }A_i({\bf x},t) = \sum_{j\neq i} a_{ij} \bar{u}_j({\bf x},t),
\end{align}\end{linenomath*}
where $a_{ij}$ are constants, which can be either positive, if population $i$ benefits from the presence of population $j$, or negative, if population $i$ seeks to avoid population $j$, and
\begin{linenomath*}\begin{align}
\label{ujbar}
\bar{u}_j({\bf x},t) = \frac{1}{|C_{\bf x}|}\int_{C_{\bf x}}u_j({\bf z},t){\rm d}{\bf z},
\end{align}\end{linenomath*}
where $C_{\bf x}$ is a small neighbourhood of ${\bf x}$, and $|C_{\bf x}|$ is the Lebesgue measure $C_{\bf x}$.  The importance of this spatial averaging will become apparent in Section \ref{sec:gen1d}.

For larger organisms (e.g. mammals, birds, reptiles etc.), individuals may be more spread-out on the landscape.  Here, presence may be advertised by one of two processes (Scenarios 2 and 3).  In Scenario 2, we model individuals as leaving marks on the landscape (e.g. urine, faeces, footprints etc.) to which individuals of the other populations respond.  Denoting  by $p_i$ the presence of marks that are foreign to population $i$, we can model this using the following differential equation (cf. \citet{lewismurray1993, moorcroftlewis2006, pottslewis2016b})
\begin{linenomath*}\begin{align}
\label{scent}
\frac{\partial p_i}{\partial t} = \sum_{j\neq i}\alpha_{ij} u_j-\mu p_i,
\end{align}\end{linenomath*}
where $\mu>0$ and $\alpha_{ij} \in {\mathbb R}$ are constants.  If $\alpha_{ij}>0$ (resp. $\alpha_{ij}<0$) then population $i$ is attracted towards (resp. repelled away from), population $j$.  In this scenario, we model $A_i({\bf x},t)$ as a spatial averaging of $p_i({\bf x},t)$ so that
\begin{linenomath*}\begin{align}
\label{Aiscent}
\mbox{\bf Scenario 2: }A_i({\bf x},t) = \bar{p}_i({\bf x},t),
\end{align}\end{linenomath*}
where $\bar{p}_i({\bf x},t)$ is defined in an analogous way to $\bar{u}_j({\bf x},t)$ in Equation (\ref{ujbar}).

Finally, Scenario 3 involves individuals remembering places where they have had recent encounters with individuals of another population, and moving in a manner consistent with a cognitive map.  We assume here that individuals within a population are able to transmit information between themselves so that they all share common information regarding the expected presence of other populations, which we denote by $k_i({\bf x},t)$ for population $i$.  This can be modelled as follows (cf. \citet{pottslewis2016a})
\begin{linenomath*}\begin{align}
\label{memory}
\frac{\partial k_i}{\partial t} = \sum_{j\neq i}\beta_{ij} u_iu_j-(\zeta+\nu u_i) k_i,
\end{align}\end{linenomath*}
where $\nu > 0, \zeta \geq 0$ and $\beta_{ij}\in {\mathbb R}$ are constants.  Here, $\beta_{ij}$ refers to the tendency for animals from population $i$ to remember a spatial location, given an interaction with an individual from population $j$, $\zeta$ is the rate of memory decay, and $\nu$ refers to the tendency for animals from population $i$ to consider a location not part of $j$'s range if individuals from $i$ visit that location without observing an individual from $j$ there.  See \citet{pottslewis2016a} more explanation of the motivation and justification for the functional form in Equation (\ref{memory}), in the context of avoidance mechanisms.

In this scenario, we model $A_i({\bf x},t)$ as a spatial averaging of $k_i({\bf x},t)$ so that
\begin{linenomath*}\begin{align}
\label{Aimemory}
\mbox{\bf Scenario 3: }A_i({\bf x},t) = \bar{k}_i({\bf x},t),
\end{align}\end{linenomath*}
where $\bar{k}_i({\bf x},t)$ is defined in an analogous way to $\bar{u}_j({\bf x},t)$ in Equation (\ref{ujbar}).  

Note the similarity between Scenarios 2 and 3 and the idea of a ``landscape of fear'', which has become increasingly popular in the empirical literature \citep{laundreetal2010}.  The landscape of fear invokes the idea that there are certain parts of space that individuals in a population tend to avoid because they perceive those areas to have a higher risk of aggressive interactions (either due to predation or competition).  The degree to which this danger is perceived across space creates a spatial distribution of fear, and animals may be modelled as advecting down the gradient of this distribution.

\section{General results in 1D}
\label{sec:gen1d}

Although our modelling framework can be defined in arbitrary dimensions, we will focus our analysis on the following 1D version of Equation (\ref{gen_model})
\begin{linenomath*}\begin{align}
\label{1Dmodel}
\frac{\partial u_i}{\partial t} = D_i\frac{\partial^2 u_i}{\partial x^2} - c_{i} \frac{\partial}{\partial x}\left(u_i \frac{\partial A_i}{\partial x} \right).
\end{align}\end{linenomath*}
We also work on a line segment, so that $x \in [0,L]$ for some $L>0$.

It is convenient for analysis to assume that, for Scenarios 2 and 3, the quantities $p_i({x},t)$ and $k_i({x},t)$ equilibriate much faster than $u_i({x},t)$, so we can make the approximations ${\partial p_i}/{\partial t}\approx 0$ and ${\partial k_i}/{\partial t}\approx 0$.  Making the further assumption that there is no memory decay ($\zeta=0$ in Equation \ref{memory}), which turns out later to be convenient for unifying the three scenarios, we have the following approximate versions of Equations (\ref{Aiscent}) and (\ref{Aimemory})
\begin{linenomath*}\begin{align}
\label{Ai23}
\mbox{\bf Scenario 2: }A_i({x},t) &\approx \sum_{j\neq i}\frac{\alpha_{ij}}{\mu} \bar{u}_j({x},t), \\
\mbox{\bf Scenario 3: }A_i({x},t) &\approx \sum_{j\neq i}\frac{\beta_{ij}}{\nu} \bar{u}_j({x},t).
\end{align}\end{linenomath*}
We non-dimensionalise our system by setting $\tilde{u}_i=Lu_i$, $\tilde{x}=x/L$, $\tilde{t}=tD_1/L^2$, $d_i=D_i/D_1$ and 
\begin{linenomath*}\begin{align}
\label{nondim}
\gamma_{ij}=
\begin{cases}
\frac{c_ia_{ij}}{LD_1},&\mbox{in Scenario 1}, \\
\frac{c_i\alpha_{ij}}{\mu LD_1},&\mbox{in Scenario 2}, \\
\frac{c_i\beta_{ij}}{\nu LD_1},&\mbox{in Scenario 3}.
\end{cases}
\end{align}\end{linenomath*}
Then, dropping the tildes over $\tilde{u}_i$, $\tilde{x}$, and $\tilde{t}$ for notational convenience, we obtain the following non-dimensional model for space use
\begin{linenomath*}\begin{align}
\label{1DmodelND}
\frac{\partial u_i}{\partial t} &= d_i\frac{\partial^2 u_i}{\partial x^2} - \frac{\partial}{\partial x}\left(u_i \sum_{j\neq i}\gamma_{ij}\frac{\partial \bar{u}_j}{\partial x} \right),
\end{align}\end{linenomath*}
where $d_1=1$, by definition.  


For simplicity, we assume that boundary conditions are periodic, so that
\begin{linenomath*}\begin{align}
\label{1dbdry} 
u_i(0,t)=u_i(1,t).
\end{align}\end{linenomath*}
With this identification in place, we can define the 1D spatial averaging kernel from Equation (\ref{ujbar}) to be $C_x=\{z \in [0,1] | (x-\delta) (\mbox{mod }1)<z<(x+\delta) (\mbox{mod }1)\}$ for $0<\delta\ll 1$.  Here, $z(\mbox{mod }1)$ is used so as to account for the periodic boundary conditions and is defined to be the unique real number $z' \in [0,1)$ such that $z-z' \in \mathbb{Z}$. Then Equation (\ref{ujbar}) becomes
\begin{linenomath*}\begin{align}
\label{ujbar1d}
\bar{u}_j = 
\frac{1}{2\delta}\int_{(x-\delta)(\mbox{mod }1)}^{(x+\delta)(\mbox{mod }1)} u_j(z,t){\rm d}z.
\end{align}\end{linenomath*}
Finally, since $u_i(x,t)$ are probability density functions of $x$, defined on the interval $x \in [0,1]$, we also have the integral condition
\begin{linenomath*}\begin{align}
\label{intcond}
\int_0^1 u_i(x,t){\rm d}x=1.
\end{align}\end{linenomath*}
This condition means that we have a unique spatially-homogeneous steady state, given by $u_i^\ast(x)=1$ for all $i \in \{1,\dots,N\}, x \in [0,1]$.  Our first task for analysis is to see whether this steady state is unstable to non-constant perturbations.

We set ${\bf w}(x,t)=(u_1-1,\dots,u_N-1)^T=(u_{1}^{(0)},\dots,u_{N}^{(0)})^T\exp(\sigma t + {\rm i}\kappa x)$, where $u_{1,0},\dots,u_{N,0}$ and $\sigma,\kappa$ are constants, and the superscript $T$ denotes matrix transpose.  By neglecting non-linear terms, Equation (\ref{1DmodelND}) becomes
\begin{linenomath*}\begin{align}
\label{lingen}
\sigma {\bf w}=\kappa^2 M(\kappa,\delta){\bf w},
\end{align}\end{linenomath*}
where $A(\kappa,\delta)=[M_{ij}(\kappa,\delta)]_{i,j}$ is a matrix with
\begin{linenomath*}\begin{align}
\label{lingen2}
M_{ij}(\kappa,\delta)=\begin{cases}
-d_i, &\mbox{if $i=j$,} \\
\gamma_{ij}\mbox{sinc}(\kappa\delta), &\mbox{otherwise,}
\end{cases}
\end{align}\end{linenomath*}
where $\mbox{sinc}(\xi)=\sin(\xi)/\xi$.  Therefore patterns form whenever there is some $\kappa$ such that there is an eigenvalue of $M(\kappa,\delta)$ with positive real part.

\begin{figure*}[ht]
	\includegraphics[width=\columnwidth]{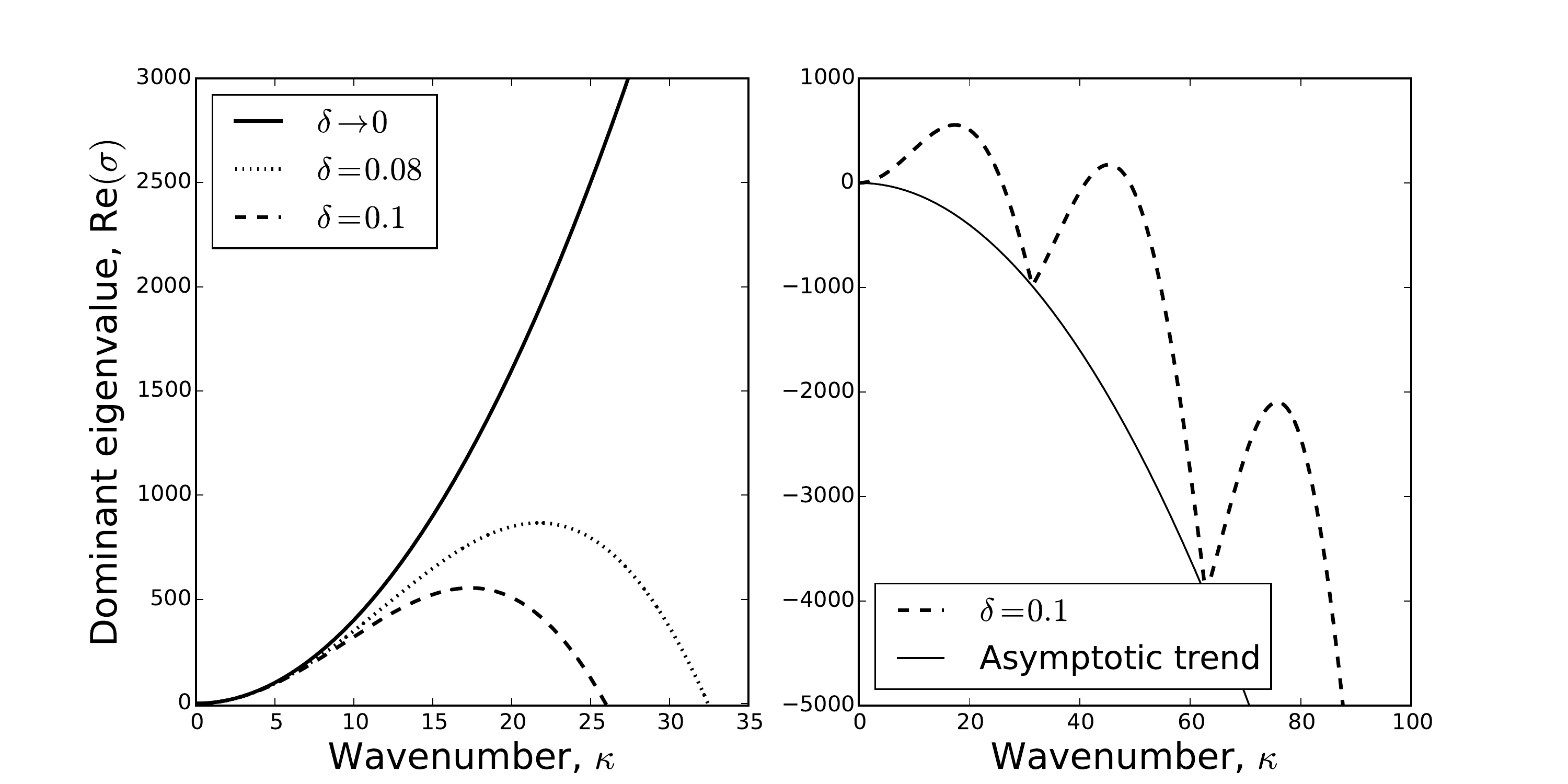}
	\caption{{\bf Example dispersion relations. } Here we give dispersion relations for the system described by Equation (\ref{1DmodelND}) with $N=2$, $d_i=1$, and $\gamma_{ij}=-5$ for all $i,j$.  In the left-hand panel, we examine three values of $\delta$, showing that, for $\delta \rightarrow 0$, the dispersion relation is monotonic, but this monotonicity is tamed by setting $\delta >0$.  In the right-hand panel, we extend the dispersion relation plot for $\delta=0.1$ to a larger range of $\kappa$ values, together with the analytically-derived asymptotic trend.}
	\label{map_disp_rel_plot}      
\end{figure*}

It is instructive to examine the limit case $\delta \rightarrow 0$.  Here
\begin{linenomath*}\begin{align}
\label{lingen3}
M_{ij}(\kappa,0)=\begin{cases}
-d_i, &\mbox{if $i=j$} \\
\gamma_{ij}, &\mbox{otherwise.}
\end{cases}
\end{align}\end{linenomath*}
so $M_{ij}(\kappa,0)$ is, in fact, independent of $\kappa$, and so we define the constant matrix $M_0=[M_{ij}(\kappa,0)]_{i,j}$.  When $\delta \rightarrow 0$, there are two cases pertinent to pattern formation:
\begin{enumerate}
\item All the eigenvalues of $M_0$ have negative real part, in which case no patterns form.
\item At least one eigenvalue $M_0$ has positive real part, in which case the dominant eigenvalue of $\kappa^2 M_0$ is an increasing function of $\kappa$.  Therefore patterns can form at arbitrarily high wavenumbers.  In other words, the pattern formation problem is ill-posed.
\end{enumerate}
The problem posed by point (2) above can often be circumvented by using a strictly positive $\delta$.  For example, Fig. \ref{map_disp_rel_plot} shows the dispersion relation (plotting the dominant eigenvalue against $\kappa$) for a simple case where $N=2$, $d_i=1$, $\gamma_{ij}=-5$ for all $i,j$, and $\delta$ is varied.  In this example, the dominant eigenvalue is real for all $\kappa$.  We see that, for $\delta \rightarrow 0$, the dispersion relation is monotonically increasing.  However, a strictly positive $\delta$ means the eigenvalues are $\kappa^2[-2 \pm 5\mbox{sinc}(\kappa\delta)]/2$, which is asymptotically $\sigma \approx -\kappa^2$ as $\kappa\rightarrow \infty$.  Hence the dominant eigenvalue is positive only for a finite range of $\kappa$-values, as long as $\delta>0$.

The fact that the pattern formation problem is ill-posed for $\delta\rightarrow 0$ suggests that classical solutions may not exist in this case.  This phenomenon is not new and has been observed in very similar systems studied by \citet{briscoeetal2002, pottslewis2016a, pottslewis2016b}.  More generally, there are various studies that deal with regularisation of such ill-posed problems in slightly different contexts using other techniques, which incorporate existence proofs (e.g. \citet{padron1998,padron2004}).  We therefore conjecture that classical solutions do exist for the system given by Equation (\ref{1DmodelND}) in the case where $\delta>0$, and the numerics detailed in this paper give evidence to support this.  However, we do not prove this conjecture here, since it is a highly non-trivial question in general, and the purpose of this paper is just to introduce the model structure and investigate possible types of patterns that could arise.  Nonetheless, it is an important subject for future research.  In the next two sections, we will examine specific cases where $N=2$ and $N=3$.

\section{The case of two interacting populations ($N=2$)}

When $N=2$, the system given by Equations (\ref{1DmodelND}, \ref{ujbar1d}, \ref{intcond}) is simple enough to categorise its linear pattern formation properties in full.  Here
\begin{linenomath*}\begin{align}
\label{M1D}
M(\kappa,\delta)=\left(
\begin{array}{cc}
-1 & \gamma_{12}\mbox{sinc}(\kappa\delta) \\
\gamma_{21}\mbox{sinc}(\kappa\delta) & -d_2
\end{array}
\right).
\end{align}\end{linenomath*}
The eigenvalues of $M(\kappa,\delta)$ are therefore
\begin{linenomath*}\begin{align}
\label{M1Dev}
\sigma(\kappa) = \frac{-(1+d_2)\pm\sqrt{(1+d_2)^2+4[\gamma_{12}\gamma_{21}\mbox{sinc}^2(\kappa\delta)-d_2]}}{2}.
\end{align}\end{linenomath*}
Notice first that if $\sigma(\kappa)$ is not real then the real part is $\mbox{Re}[\sigma(\kappa)]=-(1+d_2)/2$, which is always negative, since $d_2>0$.  Hence patterns can only form when $\sigma(\kappa)\in{\mathbb R}$, meaning that the discriminant, $\Delta=(1+d_2)^2+4[\gamma_{12}\gamma_{21}\mbox{sinc}^2(\kappa\delta)-d_2]$, must be positive.  In addition, $\sigma(\kappa)>0$ only when $\Delta>(1+d_2)^2$.  This occurs whenever $\gamma_{12}\gamma_{21}\mbox{sinc}^2(\kappa\delta) > d_2$.  Since the maximum value of $\mbox{sinc}^2(\kappa\delta)$ is 1, which is achieved at $\kappa=0$, we arrive at the following necessary criterion for pattern formation, which is also sufficient if we either drop the boundary conditions or take the $\delta \rightarrow 0$ limit
\begin{linenomath*}\begin{align}
\label{1Dpat}
\gamma_{12}\gamma_{21} > d_2.
\end{align}\end{linenomath*}
Furthermore, any patterns that do form are stationary patterns, since the eigenvalues are always real if their real part is positive.

\begin{figure*}[ht]
	\includegraphics[width=100mm]{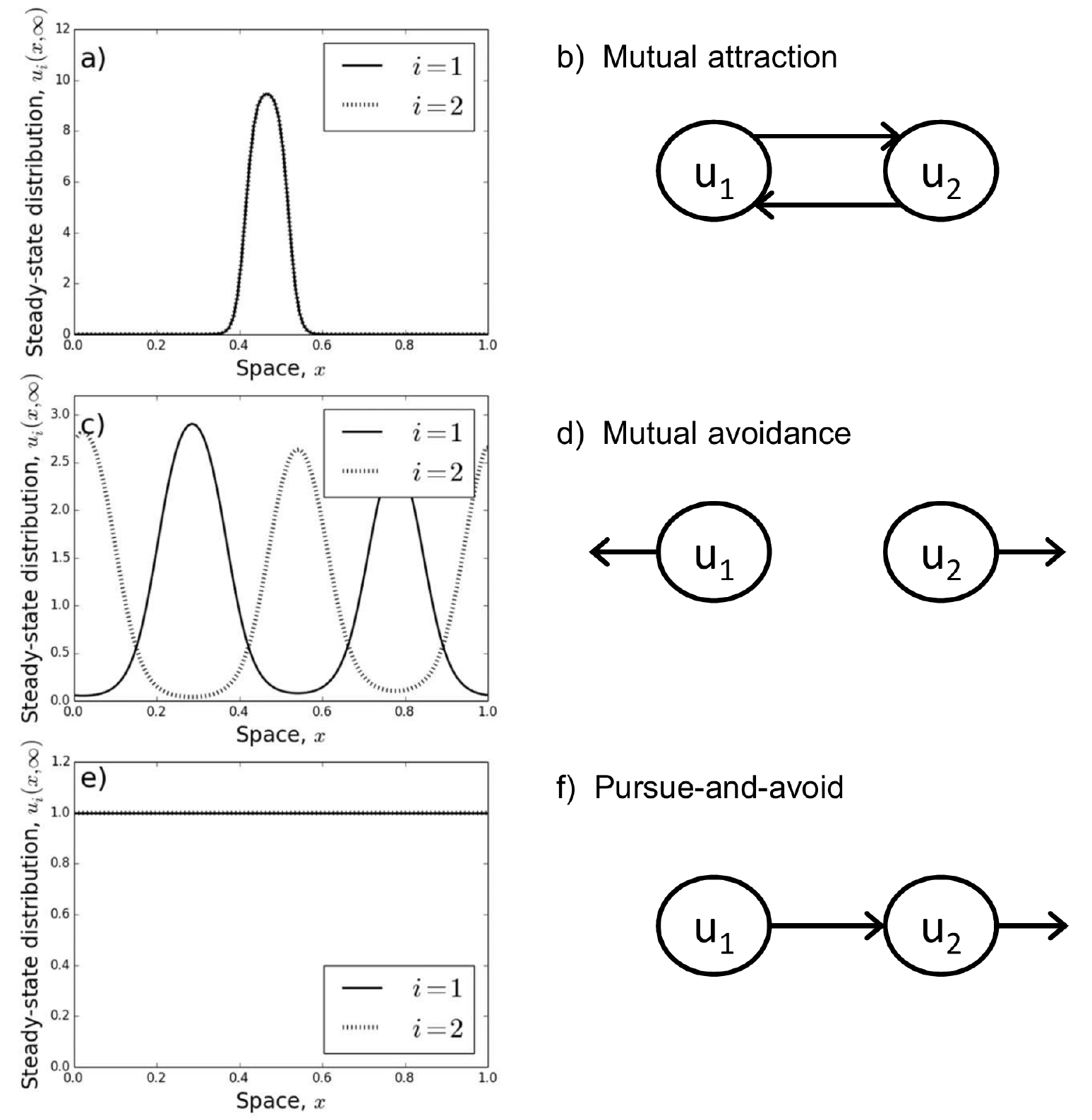}
	\caption{{\bf Dynamics for a two-species system.} Here, there are three cases: mutual attraction, mutual avoidance, and pursue-and-avoid.  Panel (a) shows the steady state of a model of mutual attraction, with $\gamma_{12}=\gamma_{21}=2$ and $\delta=0.1$, with a schematic of this situation in Panel (b).  Panel (c) is the steady state of a mutual avoidance model with $\gamma_{12}=\gamma_{21}=-2$ and $\delta=0.1$, with corresponding schematic in Panel (d).  Panel (e) is the steady state of a pursue-and-avoid model (where patterns never form) with $\gamma_{12}=2, \gamma_{21}=-2$ and $\delta=0.1$, with corresponding schematic in Panel (f).}
	\label{2SpecFig}      
\end{figure*}

There are three distinct biologically relevant situations, which correspond to different values of $\gamma_{12}$ and $\gamma_{21}$, as follows
\begin{enumerate}
\item Mutual avoidance: $\gamma_{12},\gamma_{21}<0$
\item Mutual attraction: $\gamma_{12},\gamma_{21}>0$
\item Pursue-and-avoid: $\gamma_{12}<0<\gamma_{21}$ or $\gamma_{21}<0<\gamma_{12}$
\end{enumerate}
There are also the edge cases where $\gamma_{12}=0$ or $\gamma_{21}=0$, which we will not focus on.  Notice that the `pursue-and-avoid' case cannot lead to the emergence of patterns (Fig. \ref{2SpecFig}c), as it is inconsistent with the inequality in (\ref{1Dpat}).  However, the other two situations can.  

Mutual avoidance leads to spatial segregation if Inequality (\ref{1Dpat}) is satisfied (Fig. \ref{2SpecFig}b).  Some previous models of territory formation in animal populations by the present authors have a very similar form to the mutual avoidance model here, so we refer to \citet{pottslewis2016a, pottslewis2016b} for details of this situation.  Mutual attraction leads to aggregation of both populations in a particular part of space, whose width roughly corresponds to the width of the spatial averaging kernel, $(x-\delta,x+\delta)$ (Fig. \ref{2SpecFig}a), as long as Inequality (\ref{1Dpat}) is satisfied.

The characterisation of between-population movement responses into `mutual avoidance', `mutual attraction', and `pursue-and-avoid' enables us to categorise examples of the system in Equations (\ref{1DmodelND}, \ref{ujbar1d}, \ref{intcond}) by means of a simple schematic diagram.  We construct one node for each population, ensuring that no three distinct nodes are in a straight line.  Then an arrow is added from node $i$ to node $j$ if $\gamma_{ij}>0$.  If $\gamma_{ij}<0$, an arrow is added from node $i$ in the direction anti-parallel to the line from node $i$ to node $j$.  These diagrams allow us to see quickly the qualitative relationship between the populations (see Fig. \ref{2SpecFig}d-f for the $N=2$ case and Fig. \ref{3SpecFig}b for some examples in the $N=3$ case).

\subsection{An energy functional approach to analysing patterns}

We can gain qualitative understanding of the patterns observed in Fig. \ref{2SpecFig}a-d via use of an energy functional approach, by assuming $\gamma_{1,2}=\gamma_{2,1}=\gamma$ and $d_2=1$.  In particular, this approach gives a mathematical explanation for the appearance of aggregation patterns when $\gamma>0$ and segregation patterns when $\gamma<0$.  The results rely on the assumption that, for all $i$, $u_i(x,0)> 0$ implies $u_i(x,t)> 0$ for all $t$, which can be shown by the application of a comparison theorem to Equations (\ref{1Dmodel},\ref{1dbdry}), assuming ${\partial}A_i(x)/{\partial}x$ is bounded.  
Throughout this section, our spatial co-ordinates will be defined on the quotient space $[0,1]/\{0,1\}$, which is consistent with our use of periodic boundary conditions.  

Our method makes use of the following formulation of Equation (\ref{1DmodelND})
\begin{linenomath*}\begin{align}
\label{1DmodelNDenergy}
\frac{\partial u_i}{\partial t} &= \frac{\partial}{\partial x}\left[u_i\frac{\partial}{\partial x}\left(d_i\ln(u_i) -\sum_{j\neq i}\gamma_{ij}{\mathcal K}\ast u_j \right)\right],
\end{align}\end{linenomath*}
and also the energy functional
\begin{linenomath*}\begin{align}
\label{engfunc}
E(u_1,u_2)=\int_0^1\{u_1[2\ln(u_1)-\gamma {\mathcal K} \ast u_2]+u_2[2\ln(u_2)-\gamma {\mathcal K} \ast u_1]\}{\rm d}x,
\end{align}\end{linenomath*}
where ${\mathcal K}(x)$ is a bounded function (i.e. $\left\lVert {\mathcal K}\right\rVert_{\infty} < \infty$), symmetric about $x=0$ on the domain $[0,1]/\{0,1\}$, with $\left\lVert {\mathcal K}\right\rVert_1=1$, and $\ast$ denotes the following spatial convolution
\begin{linenomath*}\begin{align}
\label{convolution}
{\mathcal K} \ast u_i(x) = \int_0^1 {\mathcal K}(x-y) u_i(y){\rm d}y.
\end{align}\end{linenomath*}
In our situation, Equation (\ref{ujbar1d}) implies that ${\mathcal K}(x)=1/(2\delta)$ for $-\delta < x < \delta\mbox{ (mod $1$)}$ and ${\mathcal K}(x)=0$ for $\delta \leq x \leq 1-\delta$.  We consider solutions $u_1(x,t)$ and $u_2(x,t)$ that are continuous functions of $x$ and $t$.

We show that the energy functional from Equation (\ref{engfunc}) decreases over time to a minimum, which represents the steady state solution of the system.  The monotonic decrease of $E$ over time is shown as follows
\begin{linenomath*}\begin{align}
\frac{\partial E}{\partial t} &= \int_0^1\left\{\frac{\partial u_1}{\partial t}[2\ln(u_1)-\gamma {\mathcal K} \ast u_2]+\frac{\partial u_2}{\partial t}[2\ln(u_2)-\gamma {\mathcal K} \ast u_1]\right\}{\rm d}x \nonumber \\
& \qquad+\int_0^1 \left[2 \frac{\partial u_1}{\partial t}+2 \frac{\partial u_2}{\partial t}-\gamma u_1 {\mathcal K} \ast \frac{\partial u_2}{\partial t}-\gamma u_2 {\mathcal K} \ast \frac{\partial u_1}{\partial t}\right]{\rm d}x\nonumber \\
&=\int_0^1\left\{2 \frac{\partial u_1}{\partial t}+2 \frac{\partial u_2}{\partial t}+\frac{\partial u_1}{\partial t}[2\ln(u_1)-2\gamma {\mathcal K}\ast u_2]+\frac{\partial u_2}{\partial t}[2\ln(u_2)-2\gamma {\mathcal K}\ast u_1]\right\}{\rm d}x \nonumber \\
&=2\int_0^1\biggl\{\frac{\partial}{\partial x}\left[u_1\frac{\partial}{\partial x}\left(\ln(u_1) -\gamma {\mathcal K}\ast u_2\right)\right][1+\ln(u_1)-\gamma {\mathcal K}\ast u_2]\nonumber \\
& \qquad+\frac{\partial}{\partial x}\left[u_2\frac{\partial}{\partial x}\left(\ln(u_2) -\gamma {\mathcal K}\ast u_1\right)\right][1+\ln(u_2)-\gamma {\mathcal K}\ast u_1]\biggr\}{\rm d}x \nonumber \\
&=2\biggl[u_1\frac{\partial}{\partial x}(\ln(u_1)-\gamma {\mathcal K}\ast u_2)(1+\ln(u_1)-\gamma {\mathcal K}\ast u_2)\nonumber \\
&\qquad\qquad +u_2\frac{\partial}{\partial x}(\ln(u_2)-\gamma {\mathcal K}\ast u_1)(1+\ln(u_2)-\gamma {\mathcal K}\ast u_1)\biggr]^1_0\nonumber \\
& \qquad -2 \int_0^1\biggl\{\left[u_1\frac{\partial}{\partial x}(\ln(u_1)-\gamma {\mathcal K}\ast u_2)\right]\frac{\partial}{\partial x}(\ln(u_1)-\gamma {\mathcal K}\ast u_2)\nonumber \\
&\qquad\qquad+\left[u_2\frac{\partial}{\partial x}(\ln(u_2)-\gamma {\mathcal K}\ast u_1)\right]\frac{\partial}{\partial x}(\ln(u_2)-\gamma {\mathcal K}\ast u_1)\biggr\}{\rm d}x \nonumber \\
&=-2 \int_0^1\biggl\{\left[u_1\frac{\partial}{\partial x}(\ln(u_1)-\gamma {\mathcal K}\ast u_2)\right]\frac{\partial}{\partial x}(\ln(u_1)-\gamma {\mathcal K}\ast u_2)\nonumber \\
&\qquad\qquad+\left[u_2\frac{\partial}{\partial x}(\ln(u_2)-\gamma {\mathcal K}\ast u_1)\right]\frac{\partial}{\partial x}(\ln(u_2)-\gamma {\mathcal K}\ast u_1)\biggr\}{\rm d}x \nonumber \\
&=-2 \int_0^1\left\{u_1\left[\frac{\partial}{\partial x}(\ln(u_1)-\gamma {\mathcal K}\ast u_2)\right]^2+u_2\left[\frac{\partial}{\partial x}(\ln(u_2)-\gamma {\mathcal K}\ast u_1)\right]^2\right\}{\rm d}x \nonumber \\
& \leq 0.
\label{dEdt}
\end{align}\end{linenomath*}
Here, the first equality uses Equation (\ref{engfunc}), the second uses the fact that $\int_0^1 f(x){\mathcal K}\ast h(x)dx=\int_0^1 h(x){\mathcal K}\ast f(x)dx$ as long as ${\mathcal K}(x)$ is symmetric about 0 in $[0,1]/\{0,1\}$, and also requires that $\gamma_{1,2}=\gamma_{2,1}=\gamma$, the third uses Equation (\ref{1DmodelNDenergy}), the fourth is integration by parts, the fifth uses the fact that $u_i(0)=u_i(1)$ and ${\mathcal K}\ast u_i(0)={\mathcal K}\ast u_i(1)$ for $i \in \{1,2\}$ (i.e. periodic boundary conditions, Equation \ref{1dbdry}), the sixth is just a rearrangement, and the inequality at the end uses the fact that $u_i(x,t)> 0$ for all $i,x,t$.  In all, Equation (\ref{dEdt}) shows that $E(u_1,u_2)$ is decreasing over time.  The following shows that $E(u_1,u_2)$ is bounded below
\begin{linenomath*}\begin{align}
E(u_1,u_2)&=2\int_0^1 [u_1\ln(u_1)+u_2\ln(u_2)]{\rm d}x-\int_0^1 [u_1 {\mathcal K}\ast u_1+u_2 {\mathcal K}\ast u_2]{\rm d}x\nonumber \\
&\geq -4{\rm e}^{-1}-\int_0^1 [u_1 {\mathcal K}\ast u_1+u_2 {\mathcal K}\ast u_2]{\rm d}x\nonumber \\
&\geq -4{\rm e}^{-1}-\left\lVert u_1\right\rVert_1\left\lVert {\mathcal K}\ast u_1\right\rVert_\infty-\left\lVert u_2\right\rVert_1\left\lVert {\mathcal K}\ast u_2\right\rVert_\infty\nonumber \\
&\geq -4{\rm e}^{-1}-\left\lVert u_1\right\rVert_1\left\lVert {\mathcal K}\right\rVert_\infty \left\lVert u_1\right\rVert_1-\left\lVert u_2\right\rVert_1\left\lVert {\mathcal K}\right\rVert_\infty \left\lVert u_2\right\rVert_1\nonumber \\
&\geq -4{\rm e}^{-1}-2\left\lVert {\mathcal K}\right\rVert_\infty.
\label{EnergyBound}
\end{align}\end{linenomath*}
Here, the first inequality uses the fact that $\mbox{inf}_{u_i\geq 0} \{u_i \ln(u_i)\}=-{\rm e}^{-1}$, the second uses H\"older's inequality, the third uses Young's inequality, and the fourth the fact that $\left\lVert u_1\right\rVert_1=1$ (Equation \ref{intcond}).  For the absence of doubt, the definition $\left\lVert f\right\rVert_p=\left(\int_0^1|f(x,t)|^p{\rm d}x\right)^{1/p}$, for $p \in [1,\infty]$, is used throughout (\ref{EnergyBound}).  Again, note that the inequality $u(x,t)>0$ is required for the sequence of inequalities in (\ref{EnergyBound}) to hold.

The inequalities in (\ref{dEdt}) and (\ref{EnergyBound}) together demonstrate that $E(u_1,u_2)$ moves towards a minimum as $t\rightarrow \infty$, which is given at the point where $\frac{\partial E}{\partial t}=0$.  The latter equation is satisfied when the following two conditions hold 
\begin{linenomath*}\begin{align}
\label{minimumEnergy1}
\ln(u_1)-\gamma {\mathcal K}\ast u_2=\eta_1, \\
\label{minimumEnergy2}
\ln(u_2)-\gamma {\mathcal K}\ast u_1=\eta_2,
\end{align}\end{linenomath*}
where $\eta_1$ and $\eta_2$ are constants.  

\begin{figure*}[ht]
	\includegraphics[width=120mm]{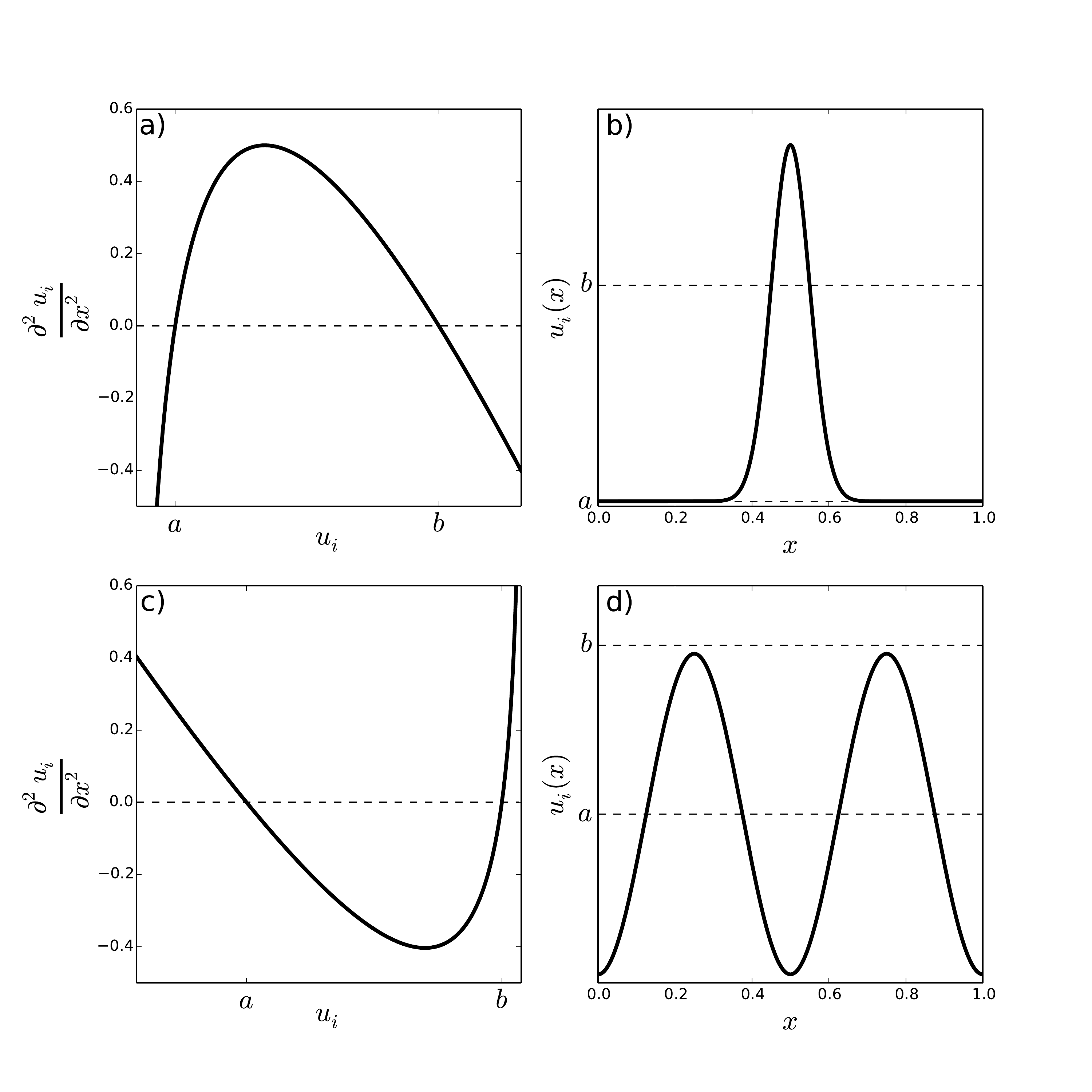}
	\caption{{\bf Understanding the patterns from Fig. \ref{2SpecFig} using energy functionals.} Panel (a) gives an example of $\frac{\partial^2 u_i}{\partial x^2}$ as a function of $u_i$ (Equation \ref{u1secderiv}) when the energy is minimised (Equations \ref{minimumEnergy1}-\ref{minimumEnergy2}) and 
the moment closure approximation from Equation (\ref{momentclosure}) is applied, for the aggregation case, $u_1 \approx u_2$.  We see that $\frac{\partial^2 u_i}{\partial x^2}$ is positive for $a<u_i<b$ and negative when $u_i<a$ or $u_i>b$.  There are various possible smooth solutions, $u_i(x,\infty)$, that satisfy this property. Panel (b) gives an example corresponding qualitatively to Fig. \ref{2SpecFig}a.  Panels (c) and (d) are analogous to (a) and (b), respectively, but for the situation where we have segregation, so $u_1 \approx 2-u_2$.  Note that Panel (d) qualitatively resembles Fig. \ref{2SpecFig}c.}
	\label{engFunc}      
\end{figure*}

Equations (\ref{minimumEnergy1}-\ref{minimumEnergy2}) can be used to give qualitative properties of the long-term distribution of the system in Equations (\ref{1DmodelND}, \ref{ujbar1d}, \ref{intcond}) for $N=2$ and $\gamma_{1,2}=\gamma_{2,1}=\gamma$.  First, by differentiating Equations (\ref{minimumEnergy1}-\ref{minimumEnergy2}) with respect to $x$, we find that 
\begin{linenomath*}\begin{align}
\label{minimumEnergy1diff}
\frac{\partial u_1}{\partial x}\frac{1}{u_1}=\gamma \frac{\partial}{\partial x}({\mathcal K}\ast u_2), \\
\label{minimumEnergy2diff}
\frac{\partial u_2}{\partial x}\frac{1}{u_2}=\gamma \frac{\partial}{\partial x}({\mathcal K}\ast u_1).
\end{align}\end{linenomath*}
Thus $\gamma>0$ implies that $\frac{\partial u_1}{\partial x}$ has the same sign as $\frac{\partial}{\partial x}({\mathcal K}\ast u_2)$ so any patterns that may form will be aggregation patterns (Fig. \ref{2SpecFig}a-b).  Furthermore, $\gamma<0$ implies that $\frac{\partial u_1}{\partial x}$ has the opposite sign to $\frac{\partial}{\partial x}({\mathcal K}\ast u_2)$ so any patterns that form will be segregation patterns (Fig. \ref{2SpecFig}c-d).

Second, by making the following moment closure approximation 
\begin{linenomath*}\begin{align}
\label{momentclosure}
{\mathcal K}\ast u_i \approx u_i+\sigma^2 \frac{\partial^2 u_i}{\partial x^2},
\end{align}\end{linenomath*}
where $\sigma^2$ is the variance of ${\mathcal K}(x)$, we can gain insight by examining the plot of $\frac{\partial^2 u_i}{\partial x^2}$ against $u_i$ in particular cases.  To give an example in the case of aggregation, if $u_1 \approx u_2$ (as in Fig. \ref{2SpecFig}a) then we have $\gamma>0$ by Equations (\ref{minimumEnergy1diff}-\ref{minimumEnergy2diff}).  Equation (\ref{minimumEnergy2}) implies 
\begin{linenomath*}\begin{align}
\label{u1secderiv}
\sigma^2 \frac{\partial^2 u_1}{\partial x^2} \approx \frac{1}{\gamma}[\ln(u_1)-\eta_2]-u_1.
\end{align}\end{linenomath*}
The right-hand side of Equation (\ref{u1secderiv}) has a unique maximum, which is above the horizontal axis as long as $\eta_2<-1-\ln(\gamma)$ (Fig. \ref{engFunc}a).  In this case, there are two numbers $a,b \in {\mathbb R}_{>0}$ such that $\frac{\partial^2 u_i}{\partial x^2}>0$ when $a<u_i<b$ and $\frac{\partial^2 u_i}{\partial x^2}<0$ for $u_i<a$ or $u_i>b$.  A possible curve that satisfies this property is given in Fig. \ref{engFunc}b, and qualitatively resembles Fig. \ref{2SpecFig}a.

To give an example in the case of segregation ($\gamma<0$), suppose that $u_1\approx 2-u_2$.  Then, by a similar argument to the $u_1 \approx u_2$ case, $\frac{\partial^2 u_i}{\partial x^2}$ has a unique minimum as long as $\eta_2<-1-\ln(-\gamma)-2\gamma$.  In this case, there are two numbers $a,b \in {\mathbb R}$ such that $\frac{\partial^2 u_i}{\partial x^2}<0$ when $a<u_i<b$ and $\frac{\partial^2 u_i}{\partial x^2}>0$ for $u_i<a$ or $u_i>b$.  A possible curve that satisfies this property is given in Fig. \ref{engFunc}d, and qualitatively resembles Fig. \ref{2SpecFig}c.

\section{The case of three interacting populations ($N=3$)}

Although the $N=2$ case only allows for stationary pattern formation (often called a {\it Turing instability} after \citet{turing1952}), for $N>2$ we can observe both stationary and oscillating patterns.  The latter arise from what is sometimes known as a {\it wave instability}, where the dominant eigenvalue of $A(\kappa,\delta)$ is not real but has positive real part, for some $\kappa$.  For $N>2$, the situation becomes too complicated for analytic expressions of the eigenvalues to give any meaningful insight (and indeed, these expressions cannot be found for $N>4$ by a classical result of Galois Theory, see \citet{stewart2015}), so we begin by examining the eigenvalues for certain example cases in the limit $\delta \rightarrow 0$.  This involves finding eigenvalues of the matrix $M_0$ given in Equation (\ref{lingen3}).

\begin{figure*}[ht]
	\includegraphics[width=\columnwidth]{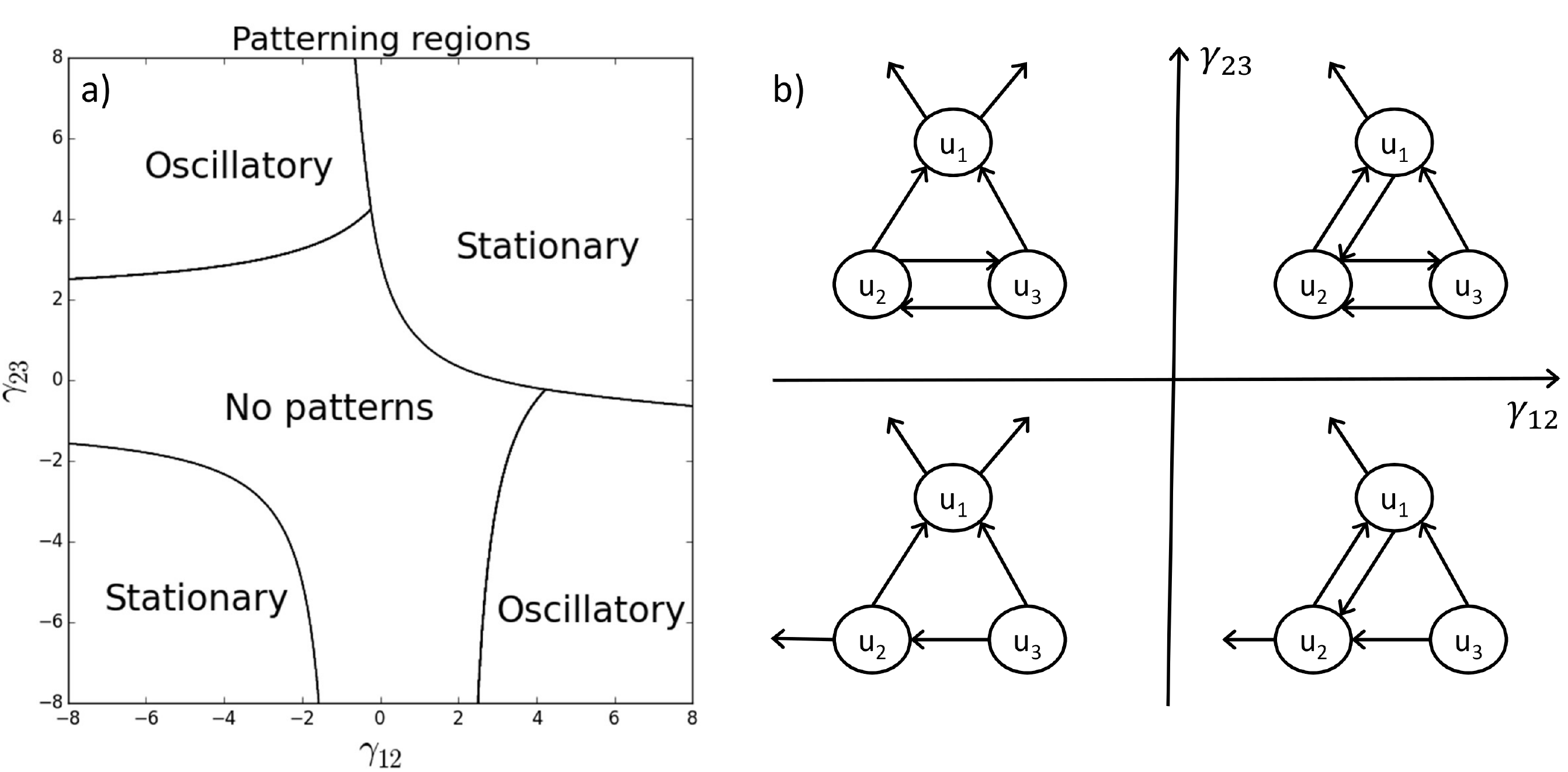}
	\caption{{\bf Dynamics for example three-species systems.} Panel (a) shows the pattern formation regions, as predicted by linear analysis, for the system in Equations (\ref{1DmodelND}, \ref{ujbar1d}, \ref{intcond}) in the case $N=3$, where $d_2=d_3=\gamma_{21}=\gamma_{31}=\gamma_{32}=1$, $\gamma_{13}=-1$, and $\gamma_{12},\gamma_{23}$ are varied.  Panel (b) shows the schematic diagrams of the systems, corresponding to the four quadrants of ($\gamma_{12},\gamma_{23}$)-space.}
	\label{3SpecFig}      
\end{figure*}

Fig. \ref{3SpecFig} gives an example of how (i) stationary patterns, (ii) oscillatory patterns, and (iii) no patterns can emerge in different regions of parameter space when $N=3$.  Here, we have fixed all the $\gamma_{ij}$ except $\gamma_{12}$ and $\gamma_{23}$.  Specifically, $d_2=d_3=\gamma_{21}=\gamma_{31}=\gamma_{32}=1$ and $\gamma_{13}=-1$.  When $\gamma_{12}<0<\gamma_{23}$ this corresponds to a mutual attraction between populations 2 and 3 with both 2 and 3 pursuing 1 in a pursue-and-avoid situation (Fig. \ref{3SpecFig}b, top-left).  When $\gamma_{12},\gamma_{23}>0$, 3 is pursuing 1 in a pursue-and-avoid, whilst 2 is mutually attracted to both 1 and 3 (Fig. \ref{3SpecFig}b, top-right).  If $\gamma_{23}<0<\gamma_{12}$, 3 is pursuing both 1 and 2 in a pursue-and-avoid, whilst 1 and 2 are mutually attracting (Fig. \ref{3SpecFig}b, bottom-right).  Finally, if $\gamma_{12},\gamma_{23}<0$ then 3 is pursuing both 1 and 2 in a pursue-and-avoid, and 2 is pursuing 1 in a pursue-and-avoid (Fig. \ref{3SpecFig}b, bottom-left).

We solved the system in Equations (\ref{1DmodelND}-\ref{intcond}) for various examples from both the stationary and oscillatory pattern regimes shown in Fig. \ref{3SpecFig}.  For this, we used periodic boundary conditions as in Equation (\ref{1dbdry}). 
We used a finite difference method, coded in \texttt{Python}, with a spatial granularity of $h=10^{-2}$ and a temporal granularity of $\tau=10^{-5}$.  Initial conditions were set to be small random fluctuations from the spatially-homogeneous steady state. 

\begin{figure*}[ht]
	\includegraphics[width=120mm]{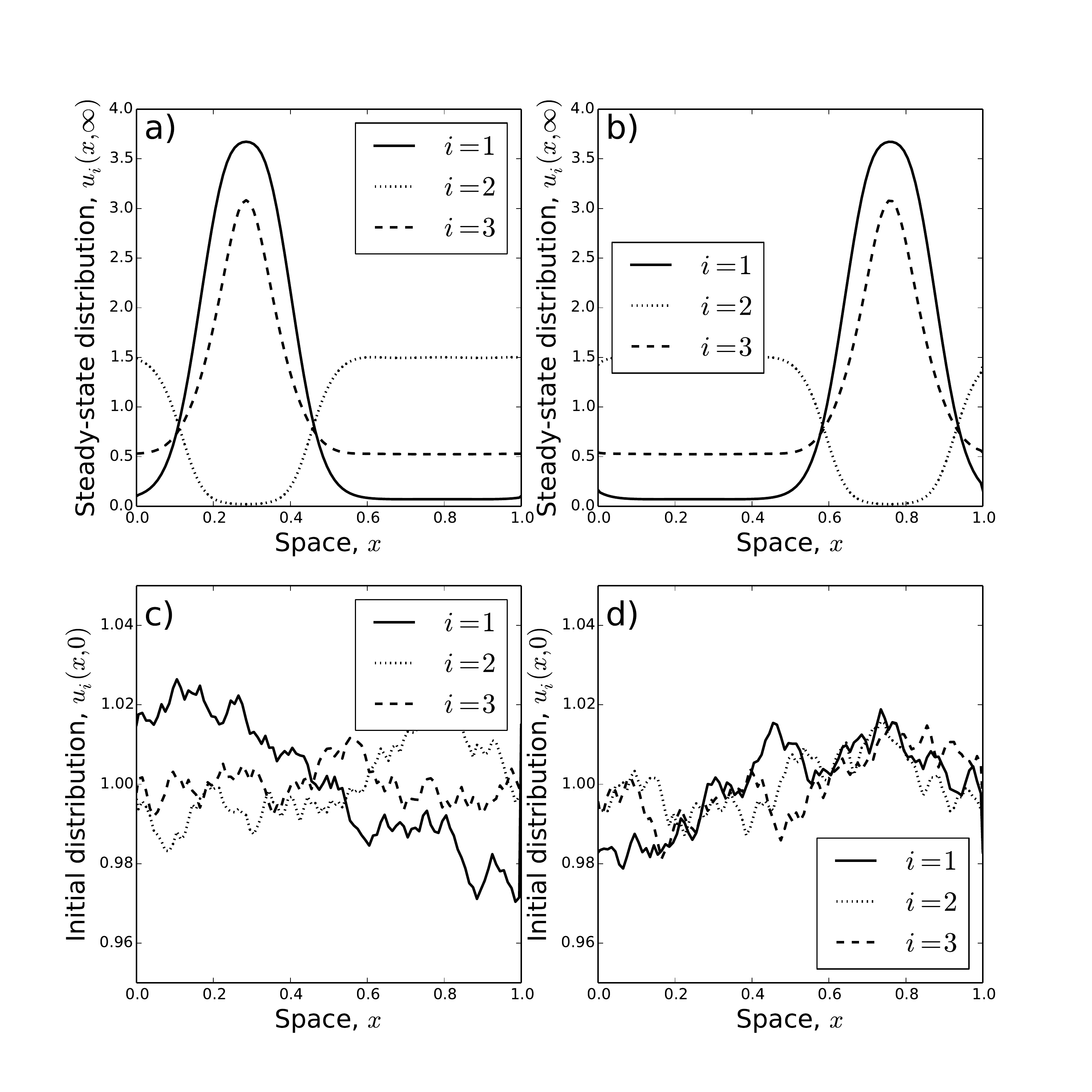}
	\caption{{\bf Example three-species systems with stationary distributions.} Panels (a) and (b) show two stable steady state distributions for the system in (\ref{1DmodelND},\ref{ujbar1d},\ref{intcond}) in the case $N=3$, where $d_2=d_3=\gamma_{21}=\gamma_{31}=\gamma_{32}=1$, $\gamma_{13}=-1$, $\gamma_{12}=\gamma_{23}=-4$, and $\delta=0.1$.  Panel (c) (resp. Panel (d)) shows the initial condition that led to the stationary distribution in Panel (a)  (resp. Panel (b)).}
	\label{3SpecStatFig}      
\end{figure*}

Stationary patterns can give rise to space partitioned into different areas for use by different populations (Fig. \ref{3SpecStatFig}, Supplementary Video SV1), with differing amounts of overlap.  Interestingly, the precise location of the segregated regions depends upon the initial conditions (compare panels (a) and (b) in Fig. \ref{3SpecStatFig}), but the rough size of the regions appears to be independent of the initial condition (at least for the parameter values we tested).  Considering the abundance of individuals as a whole (i.e. $u_1+u_2+u_3$), notice that certain regions of space emerge that contain more animals than others.  This is despite the fact that there is no environmental heterogeneity in the model.

The extent to which populations use the same parts of space depends upon the strength of attraction or repulsion.  In Fig. \ref{3SpecStatFig}a,b, the demarcation between populations 1 and 2 is quite stark, owing to the strong avoidance of population 2 by population 1 ($\gamma_{12}=-4$) and a relatively small attraction of population 2 to population 1 ($\gamma_{21}=1$).  Whereas, although population 1 seeks to avoid 3, the strength of avoidance is smaller  ($\gamma_{13}=-1$), but the attraction of population 3 to population 1 is of a similar magnitude ($\gamma_{31}=1$).  Therefore populations 1 and 3 overlap considerably.

\begin{figure*}[ht]
	\includegraphics[width=\columnwidth]{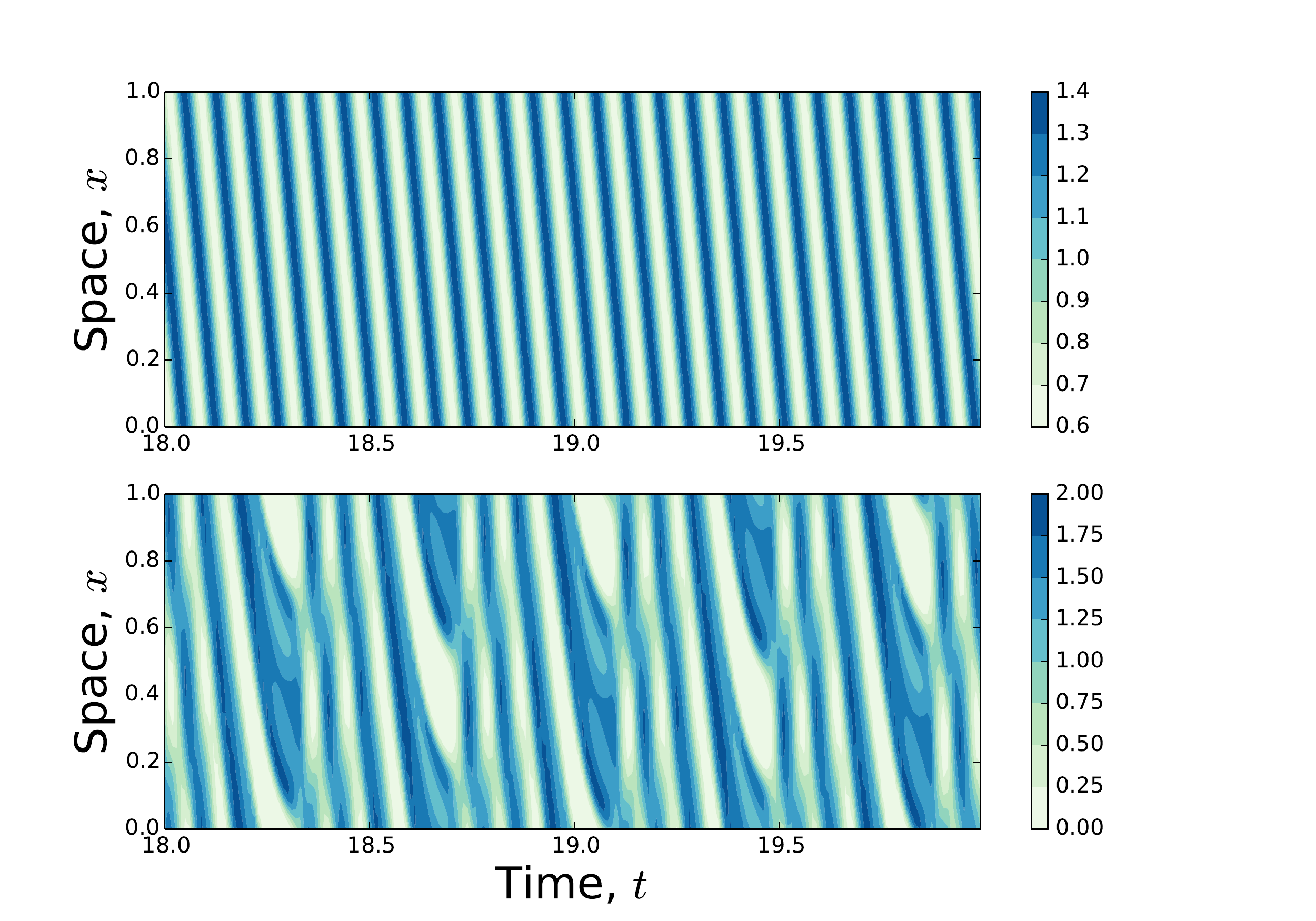}
	\caption{{\bf Example three-species systems with oscillatory distributions.} Here, we show the change in $u_1(x,t)$ over space and time for two sets of parameter values.  Both panels have parameter values identical to the fixed parameters from Fig. \ref{3SpecFig}a, additionally fixing $\gamma_{23}=-4$ and $\delta=0.1$.  Panel (a) has $\gamma_{12}=3.3$ and Panel (b) has $\gamma_{12}=4$. We started with random initial conditions and then ran the system to (dimensionless) time $t=20$.  The plots display values of $u_1(x,t)$ for $x \in [0,1]$ and $t \in [18,20]$. Plots for $t \in [14,16]$ and $t \in [16,18]$ (not shown) are very similar, indicating that the system has reached its attractor.}
	\label{3SpecOscFig}      
\end{figure*}

Oscillatory patterns can be quite complex (Supplementary Video SV2), varying from situations where there appear to be periodic oscillations (Fig. \ref{3SpecOscFig}a) to those where the periodicity is much less clear (Fig. \ref{3SpecOscFig}b).  To understand their behaviour, we use a method of numerical bifurcation analysis adapted from \citet{painterhillen2011}.  This method begins with a set of parameters in the region of no pattern formation but close to the region of oscillatory patterns.  In particular, we choose parameter values identical to the fixed values for Fig. \ref{3SpecFig}a (i.e. $d_2=d_3=\gamma_{21}=\gamma_{31}=\gamma_{32}=1$, $\gamma_{13}=-1$) and also $\gamma_{23}=-2.5$ and $\gamma_{12}=3$.  We then perform the following iterative procedure:
\begin{enumerate}
\item Solve the system numerically until $t=10$, by which time the attractor has been reached,
\item Increment $\gamma_{12}$ by a small value (we used 0.005) and set the initial conditions for the next iteration to be the final values of $u_1(x,t)$, $u_2(x,t)$, and $u_3(x,t)$ from the present numerical solution.
\end{enumerate}
This method is intended to approximate a continuous bifurcation analysis.  To analyse the resulting patterns, we focus on the value of the system for a fixed point $x=0.5$, and examine how attractor of the system $(u_1(0.5,t),u_2(0.5,t),u_3(0.5,t))$ changes as increase $\gamma_{12}$ into the region of oscillatory patterns.  

\begin{figure*}[ht]
	\includegraphics[width=\columnwidth]{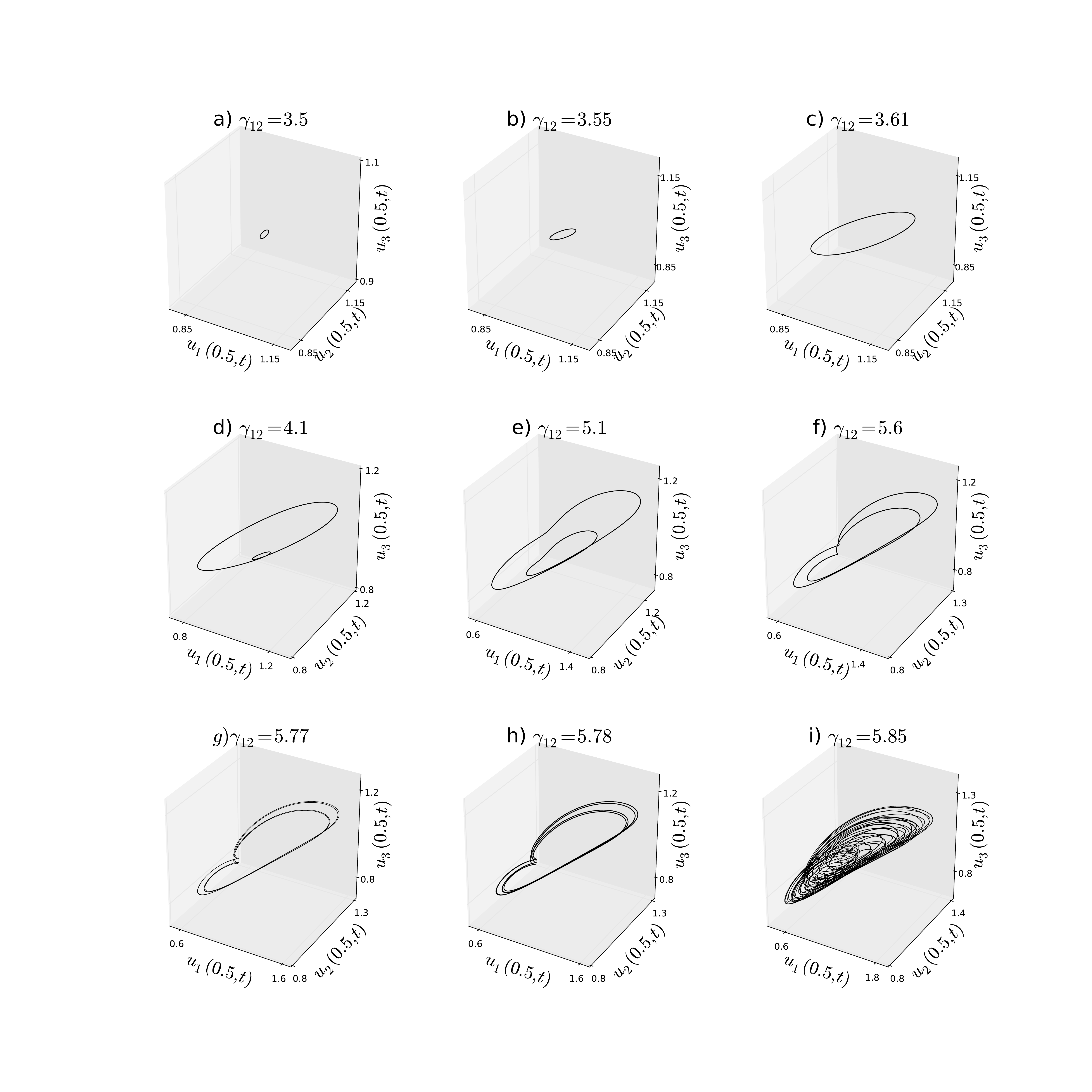}
	\caption{{\bf Numerical bifurcation analysis.} This sequence of plots shows the attractors just after the system passes through a bifurcation point from a region of no patterns to one of oscillatory patterns.  Each panel shows the locus of the point $(u_1(0.5,t),u_2(0.5,t),u_3(0.5,t))$ as time changes for a particular set of parameter values.  In all panels, $d_2=d_3=\gamma_{21}=\gamma_{31}=\gamma_{32}=1$, $\gamma_{13}=-1$, and $\gamma_{23}=-2.5$.  The value of $\gamma_{12}$ increases from panel (a) to (i) and is given in the panel title.  As $\gamma_{12}$ increases, we observe a sequence of period-doubling bifurcations leading to irregular patterns, suggestive of a chaotic system.}
	\label{map_bif_plot}      
\end{figure*}

Fig. \ref{map_bif_plot} shows these attractors for various $\gamma_{12}$-values.  First, we observe a small loop appearing just after the system goes through the bifurcation point (Fig. \ref{map_bif_plot}a). This loop then grows (Fig. \ref{map_bif_plot}b,c)	and, when $\gamma_{12}\approx 4.1$, undergoes a period-doubling bifurcation (Fig. \ref{map_bif_plot}d).	The attractor remains as a double-period loop (Fig. \ref{map_bif_plot}e,f) until $\gamma_{12}\approx 5.77$ where it doubles again (Fig. \ref{map_bif_plot}g,h). Such a sequence of period doubling is a hallmark of a chaotic system.  Indeed, as $\gamma_{12}$ is increased further, the patterns cease to have obvious period patterns (Fig. \ref{map_bif_plot}i) and gain a rather more irregular look, suggestive of chaos.

\section{Discussion}

We have used a class of diffusion-taxis systems for analysing the effect of between-population movement responses on spatial distributions of these populations.  Our models are sufficient for incorporating taxis effects due to both direct and indirect animal interactions, so are of general use for a wide range of ecological communities.  We have shown that spatial patterns in species distributions can emerge spontaneously as a result of these interactions.  What is more, these patterns may not be fixed in time, but could be in constant flux.  This brings into question the implicit assumption behind many species distribution models that the spatial distribution of a species in a fixed environment is roughly stationary over time.

Mathematically, our approach builds upon recent diffusion-taxis models of territory formation \citep{pottslewis2016a, pottslewis2016b}.  However, these latter models only consider two populations, and only in the case where there is mutual avoidance (i.e. Fig. \ref{2SpecFig}c,d).  We have shown that, when there is just one more population in the mix ($N=3$), the possible patterns that emerge can be extremely rich, incorporating stationary patterns, periodic oscillations, and irregular patterns that may be chaotic.  Although irregular and chaotic spatio-temporal patterns have been observed in spatial predator-prey systems \citep{sherratt1995ecological,sherratt1997oscillations}, this is one of the few times they have been discovered as arising from inter-population avoidance models (but see \citet[Section 8.2]{white1996model}).  These possibilities will extend to the situation of $N>3$, which is typical of most real-life ecosystems (e.g. \citet{vanaketal2013}).

The models studied here are closely related to aggregation models, which are well-studied, often with applications to cell biology in mind \citep{alt1985, mogilner1999non, topazetal2006, painteretal2015}.  In these models, populations exhibit self-attraction alongside diffusion, and are usually framed with just a single population in mind (although some incorporate more, e.g. \citet{painteretal2015, burger2018sorting}).  In contrast with our situation, this self-attraction process can enable spontaneous aggregation to occur in a single population.  Similar to our situation, in these self-attraction models it is typical to observe ill-posed problems unless some form of regularisation is in place, either through non-local terms \citep{mogilner1999non, briscoeetal2002, topazetal2006} or other means such as mixed spatio-temporal derivatives \citep{padron1998}.  

We have decided not to incorporate self-attraction into our framework.  This is both for simplicity of analysis and because the animal populations we have in mind will tend to spread on the landscape in the absence of interactions, so are well-described using diffusion as a base model \citep{okubolevin2013, lewisetal2016}.  However, in principle it is a simple extension to incorporate self-interaction into out framework, simply by dropping the $j\neq i$ restriction in Equation (\ref{1DmodelND}).  Indeed, for $N=2$, very similar models have been studied for aggregation/segregation properties \citep{burger2018sorting} and pattern formation \citep{painteretal2015}.  
In those studies, a combination of self-attraction and pursue-and-avoid can, contrary to the pure pursue-and-avoid case studied here, lead to moving spatial patterns where one aggregated population (the avoiders) leads the other one (the pursuers) in a `chase' across the landscape \citep{painter2009continuous}, a phenomenon observed in certain cell populations \citep{theveneau2013chase}. 
For $N>2$, however, we have shown that the story regarding spatial patterns can already be very rich and complicated without self-attraction, so understanding the effect of this extra complication would be a formidable exercise.

Another natural extension of our work, from a mathematical perspective, would be to add reaction terms (a.k.a. kinetics) into our model, accounting for deaths (e.g. due to predation or as a result of competition) and births, by adding a function $f_i(u_1,\dots,u_N)$ to Equation (\ref{1DmodelND}) for each $i$.  Biologically, this would change the timescale over which our model is valid, since in the present study we have explicitly set out to model timescales over which where births and deaths are negligible.  Nonetheless, this extension is worthy of discussion since the addition of such terms leads to a class of so-called cross-diffusion models, which are well-studied \citep{shigesada1979spatial, gambino2009velocity, shi2011cross, tania2012role, pottspetrovskii2017}.  The term `cross-diffusion' has been used in various guises, but the general form can incorporate both taxis terms of the type described here, as well as other terms that model various movement responses between populations.  These cross-diffusion terms can combine with the reaction terms to drive pattern formation \citep{shi2011cross, tania2012role}, as well as altering spreading speeds \citep{gambino2009velocity, girardinnadin2015}, and the outcome of competitive dynamics \citep{pottspetrovskii2017}.  The key difference between our work and traditional studies of cross-diffusion is that rich patterns form in our model despite the lack of kinetics.  As such, we separate out the effect of taxis on pattern formation from any interaction with the reaction terms.  

Our mathematical insights suggest that there is an urgent need to understand the extent to which the underlying movement processes in our model are prevalent in empirical ecosystems.  Much effort is spent in understanding species distributions \citep{manlyetal2002, araujoetal2006,jimenezetal2008}, often motivated by highly-applied questions such as understanding the effect of climate change on biodiversity loss \citep{gotellietal2015}, planning conservation efforts \citep{rodriguezetal2007, evansetal2015}, and mitigating negative effects of disease spread \citep{fatimaetal2016} and biological invasions \citep{mainalietal2015}.  Species distribution models typically seek to link the distribution of species with environmental covariates, whereas the effect of between-population movement responses is essentially ignored.  Presumably, this is because it is considered as `noise' that likely averages out over time.  In contrast, this study suggests that the patterns emerging from between-population movements may be fundamental drivers of both transient and asymptotic species distributions.  

Fortuitously, recent years have seen the development of techniques for measuring the effects of foreign populations on animal movement.  Animal bio-logging technology has become increasingly smaller, cheaper, and able to gather data at much higher frequencies than ever before \citep{wilmersetal2015, williamsetalinrev}.  Furthermore, statistical techniques have become increasingly refined to uncover the behavioural mechanisms behind animals' movement paths \citep{albertsenetal2015, avgaretal2016, michelotetal2016, pottsetal2018}.  In particular, these include inferring interactions between wild animals, both direct \citep{vanaketal2013} and mediated by environmental markers \citep{latombeetal2014, pottsetal2014b}.  

Consequently, the community of movement ecologists is in a prime position to measure between-population movement responses and seek to understand the prevalence of movement-induced spatial distribution patterns reported here.  Our hope is that the theoretical results presented here will serve as a motivating study for understanding the effect of between-population movement responses on spatial population dynamics in empirical systems, as well as highlighting the need for such studies if we are to understand accurately the drivers behind observed species distributions.

\section*{Acknowledgements}

JRP thanks the School of Mathematics and Statistics at the University of Sheffield for granting him study leave which has enabled the research presented here.  MAL gratefully acknowledges the Canada Research Chairs program and Discovery grant from the Natural Sciences and Engineering Research Council of Canada.

\bibliography{map_refs}

\begin{thebibliography}{71}
\providecommand{\natexlab}[1]{#1}
\providecommand{\url}[1]{{#1}}
\providecommand{\urlprefix}{URL }
\expandafter\ifx\csname urlstyle\endcsname\relax
  \providecommand{\doi}[1]{DOI~\discretionary{}{}{}#1}\else
  \providecommand{\doi}{DOI~\discretionary{}{}{}\begingroup
  \urlstyle{rm}\Url}\fi
\providecommand{\eprint}[2][]{\url{#2}}

\bibitem[{Adams(2001)}]{adams2001}
Adams ES (2001) Approaches to the study of territory size and shape. Ann Rev
  Ecol Syst pp 277--303

\bibitem[{Albertsen et~al.(2015)Albertsen, Whoriskey, Yurkowski, Nielsen, and
  Flemming}]{albertsenetal2015}
Albertsen CM, Whoriskey K, Yurkowski D, Nielsen A, Flemming JM (2015) Fast
  fitting of non-gaussian state-space models to animal movement data via
  template model builder. Ecology 96(10):2598--2604

\bibitem[{Alt(1985)}]{alt1985}
Alt W (1985) Degenerate diffusion equations with drift functionals modelling
  aggregation. Nonlinear Analysis: Theory, Methods \& Applications
  9(8):811--836

\bibitem[{Araujo and Guisan(2006)}]{araujoetal2006}
Araujo MB, Guisan A (2006) Five (or so) challenges for species distribution
  modelling. Journal of biogeography 33(10):1677--1688

\bibitem[{Avgar et~al.(2015)Avgar, Baker, Brown, Hagens, Kittle, Mallon,
  McGreer, Mosser, Newmaster, Patterson et~al.}]{avgaretal2015}
Avgar T, Baker JA, Brown GS, Hagens JS, Kittle AM, Mallon EE, McGreer MT,
  Mosser A, Newmaster SG, Patterson BR, et~al. (2015) Space-use behaviour of
  woodland caribou based on a cognitive movement model. Journal of Animal
  Ecology 84(4):1059--1070

\bibitem[{Avgar et~al.(2016)Avgar, Potts, Lewis, and Boyce}]{avgaretal2016}
Avgar T, Potts JR, Lewis MA, Boyce MS (2016) Integrated step selection
  analysis: bridging the gap between resource selection and animal movement.
  Methods in Ecology and Evolution 7(5):619--630

\bibitem[{B\"orger et~al.(2008)B\"orger, Dalziel, and Fryxell}]{borgeretal2008}
B\"orger L, Dalziel BD, Fryxell JM (2008) {Are there general mechanisms of
  animal home range behaviour? A review and prospects for future research}.
  Ecol Lett 11(6):637--650, \doi{10.1111/j.1461-0248.2008.01182.x},
  \urlprefix\url{http://dx.doi.org/10.1111/j.1461-0248.2008.01182.x}

\bibitem[{Briscoe et~al.(2002)Briscoe, Lewis, and Parrish}]{briscoeetal2002}
Briscoe B, Lewis M, Parrish S (2002) Home range formation in wolves due to
  scent marking. Bull Math Biol 64(2):261--284, \doi{10.1006/bulm.2001.0273},
  \urlprefix\url{http://dx.doi.org/10.1006/bulm.2001.0273}

\bibitem[{Burger et~al.(2018)Burger, Francesco, Fagioli, and
  Stevens}]{burger2018sorting}
Burger M, Francesco MD, Fagioli S, Stevens A (2018) Sorting phenomena in a
  mathematical model for two mutually attracting/repelling species. SIAM
  Journal on Mathematical Analysis 50(3):3210--3250

\bibitem[{Durrett and Levin(1994)}]{durrettlevin1994}
Durrett R, Levin S (1994) The importance of being discrete (and spatial). Theor
  Pop Biol 46(3):363--394

\bibitem[{Evans et~al.(2015)Evans, Diamond, and Kelly}]{evansetal2015}
Evans TG, Diamond SE, Kelly MW (2015) Mechanistic species distribution
  modelling as a link between physiology and conservation. Conservation
  physiology 3(1):cov056

\bibitem[{Fagan et~al.(2013)Fagan, Lewis, Auger-M\'eth\'e, Avgar, Benhamou,
  Breed, LaDage, Schl\"agel, Tang, Papastamatiou, Forester, and
  Mueller}]{faganetal2013}
Fagan WF, Lewis MA, Auger-M\'eth\'e M, Avgar T, Benhamou S, Breed G, LaDage L,
  Schl\"agel UE, Tang Ww, Papastamatiou YP, Forester J, Mueller T (2013)
  Spatial memory and animal movement. Ecol Lett 16(10):1316--1329,
  \doi{10.1111/ele.12165}, \urlprefix\url{http://dx.doi.org/10.1111/ele.12165}

\bibitem[{Fatima et~al.(2016)Fatima, Atif, Rasheed, Zaidi, and
  Hussain}]{fatimaetal2016}
Fatima SH, Atif S, Rasheed SB, Zaidi F, Hussain E (2016) Species distribution
  modelling of aedes aegypti in two dengue-endemic regions of pakistan.
  Tropical Medicine \& International Health 21(3):427--436

\bibitem[{Fleming et~al.(2015)Fleming, Fagan, Mueller, Olson, Leimgruber, and
  Calabrese}]{flemingetal2015}
Fleming CH, Fagan WF, Mueller T, Olson KA, Leimgruber P, Calabrese JM (2015)
  Rigorous home range estimation with movement data: a new autocorrelated
  kernel density estimator. Ecology 96(5):1182--1188

\bibitem[{Gallagher et~al.(2017)Gallagher, Creel, Wilson, and
  Cooke}]{gallagheretal2017}
Gallagher AJ, Creel S, Wilson RP, Cooke SJ (2017) Energy landscapes and the
  landscape of fear. Trends in Ecology \& Evolution 32(2):88--96

\bibitem[{Gambino et~al.(2009)Gambino, Lombardo, and
  Sammartino}]{gambino2009velocity}
Gambino G, Lombardo MC, Sammartino M (2009) A velocity--diffusion method for a
  lotka--volterra system with nonlinear cross and self-diffusion. Applied
  Numerical Mathematics 59(5):1059--1074

\bibitem[{Girardin and Nadin(2015)}]{girardinnadin2015}
Girardin L, Nadin G (2015) Travelling waves for diffusive and strongly
  competitive systems: relative motility and invasion speed. European Journal
  of Applied Mathematics 26(4):521--534

\bibitem[{Giuggioli et~al.(2013)Giuggioli, Potts, Rubenstein, and
  Levin}]{giuggiolietal2013}
Giuggioli L, Potts JR, Rubenstein DI, Levin SA (2013) Stigmergy, collective
  actions, and animal social spacing. Proceedings of the National Academy of
  Sciences p 201307071

\bibitem[{Gotelli and Stanton-Geddes(2015)}]{gotellietal2015}
Gotelli NJ, Stanton-Geddes J (2015) Climate change, genetic markers and species
  distribution modelling. Journal of Biogeography 42(9):1577--1585

\bibitem[{Hastings(1980)}]{hastings1980}
Hastings A (1980) Disturbance, coexistence, history, and competition for space.
  Theoretical population biology 18(3):363--373

\bibitem[{Hastings et~al.(2005)Hastings, Cuddington, Davies, Dugaw, Elmendorf,
  Freestone, Harrison, Holland, Lambrinos, Malvadkar
  et~al.}]{hastings2005spatial}
Hastings A, Cuddington K, Davies KF, Dugaw CJ, Elmendorf S, Freestone A,
  Harrison S, Holland M, Lambrinos J, Malvadkar U, et~al. (2005) The spatial
  spread of invasions: new developments in theory and evidence. Ecology Letters
  8(1):91--101

\bibitem[{Hays et~al.(2016)Hays, Ferreira, Sequeira, Meekan, Duarte, Bailey,
  Bailleul, Bowen, Caley, Costa et~al.}]{haysetal2016}
Hays GC, Ferreira LC, Sequeira AM, Meekan MG, Duarte CM, Bailey H, Bailleul F,
  Bowen WD, Caley MJ, Costa DP, et~al. (2016) Key questions in marine megafauna
  movement ecology. Trends in Ecology \& Evolution 31(6):463--475

\bibitem[{Hooten et~al.(2017)Hooten, Johnson, McClintock, and
  Morales}]{hootenetal2017}
Hooten MB, Johnson DS, McClintock BT, Morales JM (2017) Animal movement:
  statistical models for telemetry data. CRC Press

\bibitem[{Jim{\'e}nez-Valverde et~al.(2008)Jim{\'e}nez-Valverde, Lobo, and
  Hortal}]{jimenezetal2008}
Jim{\'e}nez-Valverde A, Lobo JM, Hortal J (2008) Not as good as they seem: the
  importance of concepts in species distribution modelling. Diversity and
  distributions 14(6):885--890

\bibitem[{Kareiva and Odell(1987)}]{kareivaodell1987}
Kareiva P, Odell G (1987) Swarms of predators exhibit ``prey taxis'' if
  individual predators use area-restricted search. The American Naturalist
  130(2):233--270

\bibitem[{Kays et~al.(2015)Kays, Crofoot, Jetz, and Wikelski}]{kaysetal2015}
Kays R, Crofoot MC, Jetz W, Wikelski M (2015) Terrestrial animal tracking as an
  eye on life and planet. Science 348(6240):aaa2478

\bibitem[{Kneitel and Chase(2004)}]{kneitelchase2004}
Kneitel JM, Chase JM (2004) Trade-offs in community ecology: linking spatial
  scales and species coexistence. Ecology Letters 7(1):69--80

\bibitem[{Latombe et~al.(2014)Latombe, Fortin, and Parrott}]{latombeetal2014}
Latombe G, Fortin D, Parrott L (2014) Spatio-temporal dynamics in the response
  of woodland caribou and moose to the passage of grey wolf. Journal of Animal
  Ecology 83(1):185--198

\bibitem[{Laundr{\'e} et~al.(2010)Laundr{\'e}, Hern{\'a}ndez, and
  Ripple}]{laundreetal2010}
Laundr{\'e} JW, Hern{\'a}ndez L, Ripple WJ (2010) The landscape of fear:
  ecological implications of being afraid. Open Ecology Journal 3:1--7

\bibitem[{Lee et~al.(2009)Lee, Hillen, and Lewis}]{leeetal2009}
Lee J, Hillen T, Lewis M (2009) Pattern formation in prey-taxis systems.
  Journal of biological dynamics 3(6):551--573

\bibitem[{Lewis and Moorcroft(2006)}]{moorcroftlewis2006}
Lewis M, Moorcroft P (2006) Mechanistic Home Range Analysis. Princeton
  University Press, Princeton University Press, Princeton, USA

\bibitem[{Lewis and Murray(1993)}]{lewismurray1993}
Lewis MA, Murray JD (1993) Modelling territoriality and wolf-deer interactions.
  Nature 366:738--740

\bibitem[{Lewis et~al.(2016)Lewis, Petrovskii, and Potts}]{lewisetal2016}
Lewis MA, Petrovskii SV, Potts JR (2016) The mathematics behind biological
  invasions, vol~44. Springer

\bibitem[{Lugo and McKane(2008)}]{lugomckane2008}
Lugo CA, McKane AJ (2008) Quasicycles in a spatial predator-prey model.
  Physical Review E 78(5):051911

\bibitem[{Mainali et~al.(2015)Mainali, Warren, Dhileepan, McConnachie,
  Strathie, Hassan, Karki, Shrestha, and Parmesan}]{mainalietal2015}
Mainali KP, Warren DL, Dhileepan K, McConnachie A, Strathie L, Hassan G, Karki
  D, Shrestha BB, Parmesan C (2015) Projecting future expansion of invasive
  species: comparing and improving methodologies for species distribution
  modeling. Global change biology 21(12):4464--4480

\bibitem[{Manly et~al.(2002)Manly, McDonald, Thomas, McDonald, and
  Erikson}]{manlyetal2002}
Manly B, McDonald L, Thomas D, McDonald T, Erikson W (2002) Resource selection
  by animals: statistical design and analysis for field studies. Elsevier
  Academic Press, Chapman and Hall, New York, New York, USA

\bibitem[{Michelot et~al.(2016)Michelot, Langrock, and
  Patterson}]{michelotetal2016}
Michelot T, Langrock R, Patterson TA (2016) movehmm: An r package for the
  statistical modelling of animal movement data using hidden markov models.
  Methods in Ecology and Evolution 7(11):1308--1315

\bibitem[{Mogilner and Edelstein-Keshet(1999)}]{mogilner1999non}
Mogilner A, Edelstein-Keshet L (1999) A non-local model for a swarm. Journal of
  Mathematical Biology 38(6):534--570

\bibitem[{Murray(2003)}]{murray2003}
Murray JD (2003) Mathematical biology II: Spatial models and biomedical
  applications. Springer-Verlag New York Incorporated New York

\bibitem[{Murrell and Law(2003)}]{murrelllaw2003}
Murrell DJ, Law R (2003) Heteromyopia and the spatial coexistence of similar
  competitors. Ecology letters 6(1):48--59

\bibitem[{Nathan and Giuggioli(2013)}]{nathangiuggioli2013}
Nathan R, Giuggioli L (2013) A milestone for movement ecology research.
  Movement Ecology 1(1)

\bibitem[{Nathan et~al.(2008)Nathan, Getz, Revilla, Holyoak, Kadmon, Saltz, and
  Smouse}]{nathanetal2008}
Nathan R, Getz WM, Revilla E, Holyoak M, Kadmon R, Saltz D, Smouse PE (2008) A
  movement ecology paradigm for unifying organismal movement research.
  Proceedings of the National Academy of Sciences 105(49):19052--19059,
  \doi{10.1073/pnas.0800375105},
  \urlprefix\url{http://www.pnas.org/content/105/49/19052.abstract},
  \eprint{http://www.pnas.org/content/105/49/19052.full.pdf+html}

\bibitem[{Okubo and Levin(2013)}]{okubolevin2013}
Okubo A, Levin SA (2013) Diffusion and ecological problems: modern
  perspectives, vol~14. Springer Science \& Business Media

\bibitem[{Padr{\'o}n(1998)}]{padron1998}
Padr{\'o}n V (1998) Sobolev regularization of a nonlinear ill-posed parabolic
  problem as a model for aggregating populations. Communications in partial
  differential equations 23(3-4):457--486

\bibitem[{Padr{\'o}n(2004)}]{padron2004}
Padr{\'o}n V (2004) Effect of aggregation on population recovery modeled by a
  forward-backward pseudoparabolic equation. Transactions of the American
  Mathematical Society 356(7):2739--2756

\bibitem[{Painter et~al.(2015)Painter, Bloomfield, Sherratt, and
  Gerisch}]{painteretal2015}
Painter K, Bloomfield J, Sherratt J, Gerisch A (2015) A nonlocal model for
  contact attraction and repulsion in heterogeneous cell populations. Bulletin
  of mathematical biology 77(6):1132--1165

\bibitem[{Painter(2009)}]{painter2009continuous}
Painter KJ (2009) Continuous models for cell migration in tissues and
  applications to cell sorting via differential chemotaxis. Bulletin of
  Mathematical Biology 71(5):1117

\bibitem[{Painter and Hillen(2011)}]{painterhillen2011}
Painter KJ, Hillen T (2011) Spatio-temporal chaos in a chemotaxis model.
  Physica D 240:363–375

\bibitem[{Pascual(1993)}]{pascual1993}
Pascual M (1993) Diffusion-induced chaos in a spatial predator--prey system.
  Proc R Soc Lond B 251(1330):1--7

\bibitem[{Petrovskii et~al.(2002)Petrovskii, Morozov, and
  Venturino}]{petrovskiietal2002}
Petrovskii SV, Morozov AY, Venturino E (2002) Allee effect makes possible
  patchy invasion in a predator--prey system. Ecology Letters 5(3):345--352

\bibitem[{Potts and Lewis(2014)}]{pottslewis2014}
Potts JR, Lewis MA (2014) How do animal territories form and change? lessons
  from 20 years of mechanistic modelling. Proc Roy Soc B 281(1784):20140231

\bibitem[{Potts and Lewis(2016{\natexlab{a}})}]{pottslewis2016a}
Potts JR, Lewis MA (2016{\natexlab{a}}) How memory of direct animal
  interactions can lead to territorial pattern formation. J Roy Soc Interface

\bibitem[{Potts and Lewis(2016{\natexlab{b}})}]{pottslewis2016b}
Potts JR, Lewis MA (2016{\natexlab{b}}) Territorial pattern formation in the
  absence of an attractive potential. J Math Biol 72(1-2):25--46

\bibitem[{Potts and Petrovskii(2017)}]{pottspetrovskii2017}
Potts JR, Petrovskii SV (2017) Fortune favours the brave: Movement responses
  shape demographic dynamics in strongly competing populations. Journal of
  Theoretical Biology 420:190--199

\bibitem[{Potts et~al.(2014)Potts, Mokross, and Lewis}]{pottsetal2014b}
Potts JR, Mokross K, Lewis MA (2014) A unifying framework for quantifying the
  nature of animal interactions. Journal of The Royal Society Interface
  11(96):20140333

\bibitem[{Potts et~al.(2018)Potts, B{\"o}rger, Scantlebury, Bennett, Alagaili,
  and Wilson}]{pottsetal2018}
Potts JR, B{\"o}rger L, Scantlebury DM, Bennett NC, Alagaili A, Wilson RP
  (2018) Finding turning-points in ultra-high-resolution animal movement data.
  Methods in Ecology and Evolution 9(10):2091--2101

\bibitem[{Rodr{\'\i}guez et~al.(2007)Rodr{\'\i}guez, Brotons, Bustamante, and
  Seoane}]{rodriguezetal2007}
Rodr{\'\i}guez JP, Brotons L, Bustamante J, Seoane J (2007) The application of
  predictive modelling of species distribution to biodiversity conservation.
  Diversity and Distributions 13(3):243--251

\bibitem[{Sherratt et~al.(1995)Sherratt, Lewis, and
  Fowler}]{sherratt1995ecological}
Sherratt JA, Lewis MA, Fowler AC (1995) Ecological chaos in the wake of
  invasion. Proceedings of the National Academy of Sciences 92(7):2524--2528

\bibitem[{Sherratt et~al.(1997)Sherratt, Eagan, and
  Lewis}]{sherratt1997oscillations}
Sherratt JA, Eagan BT, Lewis MA (1997) Oscillations and chaos behind
  predator--prey invasion: mathematical artifact or ecological reality?
  Philosophical transactions of the Royal Society of London Series B:
  Biological Sciences 352(1349):21--38

\bibitem[{Shi et~al.(2011)Shi, Xie, and Little}]{shi2011cross}
Shi J, Xie Z, Little K (2011) Cross-diffusion induced instability and stability
  in reaction-diffusion systems. Journal of Applied Analysis and Computation
  1(1):95--119

\bibitem[{Shigesada et~al.(1979)Shigesada, Kawasaki, and
  Teramoto}]{shigesada1979spatial}
Shigesada N, Kawasaki K, Teramoto E (1979) Spatial segregation of interacting
  species. Journal of theoretical biology 79(1):83--99

\bibitem[{Stewart(2015)}]{stewart2015}
Stewart IN (2015) Galois theory. CRC Press

\bibitem[{Sun et~al.(2012)Sun, Zhang, Song, Jin, and Li}]{sunetal2012}
Sun GQ, Zhang J, Song LP, Jin Z, Li BL (2012) Pattern formation of a spatial
  predator--prey system. Applied Mathematics and Computation
  218(22):11151--11162

\bibitem[{Tania et~al.(2012)Tania, Vanderlei, Heath, and
  Edelstein-Keshet}]{tania2012role}
Tania N, Vanderlei B, Heath JP, Edelstein-Keshet L (2012) Role of social
  interactions in dynamic patterns of resource patches and forager aggregation.
  Proceedings of the National Academy of Sciences 109(28):11228--11233

\bibitem[{Theveneau et~al.(2013)Theveneau, Steventon, Scarpa, Garcia, Trepat,
  Streit, and Mayor}]{theveneau2013chase}
Theveneau E, Steventon B, Scarpa E, Garcia S, Trepat X, Streit A, Mayor R
  (2013) Chase-and-run between adjacent cell populations promotes directional
  collective migration. Nature cell biology 15(7):763

\bibitem[{Topaz et~al.(2006)Topaz, Bertozzi, and Lewis}]{topazetal2006}
Topaz CM, Bertozzi AL, Lewis MA (2006) A nonlocal continuum model for
  biological aggregation. Bulletin of mathematical biology 68(7):1601

\bibitem[{Turing(1952)}]{turing1952}
Turing AM (1952) The chemical basis of morphogenesis. Phil Trans R Soc Lond B
  237(641):37--72

\bibitem[{Vanak et~al.(2013)Vanak, Fortin, Thakera, Ogdene, Owena, Greatwood,
  and Slotow}]{vanaketal2013}
Vanak A, Fortin D, Thakera M, Ogdene M, Owena C, Greatwood S, Slotow R (2013)
  Moving to stay in place - behavioral mechanisms for coexistence of african
  large carnivores. Ecology 94:2619--2631

\bibitem[{White et~al.(1996)White, Lewis, and Murray}]{white1996model}
White K, Lewis M, Murray J (1996) A model for wolf-pack territory formation and
  maintenance. Journal of Theoretical Biology 178(1):29--43

\bibitem[{Williams et~al.(in review)Williams, Taylor, Benhamou, Bijleveld,
  Clay, de~Grissac, Dem\v{s}ar, English, Franconi, G\'omez-Laich, Griffiths,
  Kay, Morales, Potts, Rogerson, Rutz, Spelt, Trevail, Wilson, and
  B\"orger}]{williamsetalinrev}
Williams HJ, Taylor LA, Benhamou S, Bijleveld AI, Clay TA, de~Grissac S,
  Dem\v{s}ar U, English HM, Franconi N, G\'omez-Laich A, Griffiths RC, Kay WP,
  Morales JM, Potts JR, Rogerson KF, Rutz C, Spelt A, Trevail AM, Wilson RP,
  B\"orger L (in review) Optimising the use of bio-loggers for movement ecology
  research. Journal of Animal Ecology

\bibitem[{Wilmers et~al.(2015)Wilmers, Nickel, Bryce, Smith, Wheat, and
  Yovovich}]{wilmersetal2015}
Wilmers CC, Nickel B, Bryce CM, Smith JA, Wheat RE, Yovovich V (2015) The
  golden age of bio-logging: how animal-borne sensors are advancing the
  frontiers of ecology. Ecology 96(7):1741--1753

\end{thebibliography}

\end{document}